\documentstyle[12pt,epsf]{article} 

\def\singlespace{\baselineskip=14pt}

\def\singlespace{%
    \lineskip                .15ex
    \baselineskip            3.0ex
   \lineskiplimit              0ex
   \parskip                0.60ex plus .30ex minus .15ex
   }%

\def\linebreak{\hfil\break}

\def\
{\hfil\linebreak}

\newenvironment{references}{
   \begin{list}{}{
      \labelwidth0cm
      \leftmargin3em
      \listparindent-3em
      \itemindent-3em
      \itemsep-1ex
   }
}{\end{list}}
\def\spose#1{\hbox to 0pt{#1\hss}}

\def\Mika{Miko{\l}ajewska}

\def\etal{{et al}. }

\def\MSUN{\rm M_{\odot}}
\def\RSUN{\rm R_{\odot}}

\def\MDOT{\dot{M}}
\def\GMCM2{\rm gm\,cm^{-2}}
\def\GMCM3{\rm gm\,cm^{-3}}

\def\kms{\ifmmode {\, \rm km \, s^{-1}}\else {$\, \rm km \, s^{-1}$}\fi }

\def\ang{\rm \AA}

\def\degree{\ifmmode {^\circ}\else {$^\circ$}\fi}
\def\rstar{\ifmmode {R_{\star}}\else $R_{\star}$\fi}
\def\bstar{\ifmmode {B_{\star}}\else $B_{\star}$\fi}
\def\pstar{\ifmmode {P_{\star}}\else $P_{\star}$\fi}
\def\rsun{\ifmmode {\rm R_{\odot}}\else $\rm R_{\odot}$\fi}
\def\RSUN{\ifmmode {\rm R_{\odot}}\else $\rm R_{\odot}$\fi}
\def\rsunsq{\ifmmode {\rm R_{\odot}^2}\else $\rm R_{\odot}^2$\fi}
\def\mstar{\ifmmode {M_{\star}}\else $M_{\star}$\fi}
\def\lstar{\ifmmode {L_{\star}}\else $L_{\star}$\fi}
\def\tstar{\ifmmode {T_{\star}}\else $T_{\star}$\fi}
\def\msun{\ifmmode {\rm M_{\odot}}\else $\rm M_{\odot}$\fi}
\def\MSUN{\ifmmode {\rm M_{\odot}}\else $\rm M_{\odot}$\fi}
\def\msunyr{\ifmmode {\rm M_{\odot}\,yr^{-1}}\else $\rm M_{\odot}\,yr^{-1}$\fi}
\def\mdot{\ifmmode {\dot{M}}\else $\dot{M}$\fi}
\def\lsun{\ifmmode {\rm L_{\odot}}\else $\rm L_{\odot}$\fi}
\def\LSUN{\ifmmode {\rm L_{\odot}}\else $\rm L_{\odot}$\fi}
\def\lbol{\ifmmode {L_{bol}}\else $L_{bol}$\fi}
\def\teff{\ifmmode {T_{eff}}\else $T_{eff}$\fi}
\def\ne{\ifmmode {n_{e}}\else $n_{e}$\fi}
\def\te{\ifmmode {T_{e}}\else $T_{e}$\fi}
\def\rc{\ifmmode {R_{c}}\else $R_{c}$\fi}
\def\cm3{\ifmmode {\rm cm^{-3}}\else $\rm cm^{-3}$\fi}
\def\emm{\ifmmode {n_e^2 V}\else $n_e^2 V$\fi}
\def\gcm3{\ifmmode {\rm g~cm^{-3}}\else $\rm g~cm^{-3}$\fi}
\def\ergg{\ifmmode {\rm erg~g^{-1}}\else $\rm erg~g^{-1}$\fi}
\def\ergs{\ifmmode {\rm erg~s^{-1}}\else $\rm erg~s^{-1}$\fi}
\def\ecs{\ifmmode {\rm erg~cm^{-2}~s^{-1}}\else $\rm erg~cm^{-2}~s^{-1}$\fi}
\def\mum{\ifmmode {\rm \mu {\rm m}}\else $\rm \mu {\rm m}$\fi}
\def\nh3{\ifmmode {\rm NH_3}\else $\rm NH_3$\fi}
\def\arcsec{\ifmmode ^{\prime \prime}\else $^{\prime \prime}$\fi}

\def\inch{\ifmmode ^{\prime \prime}\else $^{\prime \prime}$\fi}
\def\arcmin{\ifmmode ^{\prime}\else $^{\prime}$\fi}

\def\lfl{\ifmmode {\lambda F_{\lambda}}\else $\lambda F_{\lambda}$\fi}
\def\lFl{\ifmmode {\lambda F_{\lambda}}\else $\lambda F_{\lambda}$\fi}
\newbox\grsign \setbox\grsign=\hbox{$>$} \newdimen\grdimen \grdimen=\ht\grsign
\newbox\simlessbox \newbox\simgreatbox
\setbox\simgreatbox=\hbox{\raise.5ex\hbox{$>$}\llap
     {\lower.5ex\hbox{$\sim$}}}\ht1=\grdimen\dp1=0pt
\setbox\simlessbox=\hbox{\raise.5ex\hbox{$<$}\llap
     {\lower.5ex\hbox{$\sim$}}}\ht2=\grdimen\dp2=0pt
\def\simgreat{\mathrel{\copy\simgreatbox}}
\def\simless{\mathrel{\copy\simlessbox}}
\begin{document}
\newcommand{\Appendix}[1]{
\appendix
\section*{Appendix #1}
\setcounter{equation}{0}
\renewcommand{\theequation}{{\rm #1}\arabic{equation}}}
{\noindent}
\pagestyle{empty}
.

\vskip 7ex
\centerline{\Large {\bf Illumination in symbiotic binary stars:}}

\centerline{\Large  {\bf Non-LTE photoionization models. }}

\centerline{\  {\bf II.  Wind case.}}
\vskip 8ex

\singlespace
\centerline{Daniel Proga\footnote{also 
Harvard-Smithsonian Center for Astrophysics, 60 Garden Street, 
Cambridge, MA 02138}}
\centerline{Imperial College of Science, Technology and Medicine, }
\centerline{Prince Consort Road, London SW7 2BZ, U.K.}
\vskip 3ex

\centerline{Scott J. Kenyon, and John C. Raymond} 
\centerline{Harvard-Smithsonian Center for Astrophysics}
\centerline{60 Garden Street, Cambridge, MA 02138}



\vskip 4ex
\centerline{E-mail: d.proga@ic.ac.uk, skenyon@cfa.harvard.edu,}
\centerline{jraymond@cfa.harvard.edu}
\vskip 23ex

\vskip 10ex
\centerline{To appear in the}
\centerline{\it Astrophysical Journal}
\vfill
\eject

\pagestyle{plain}

\vskip .6cm
\centerline{\bf ABSTRACT}
\vskip 0.4cm

We describe a non-LTE photoionization code to calculate the 
wind structure and emergent spectrum of a red giant wind 
illuminated by the hot component of a symbiotic binary system.
We consider spherically symmetric winds with several different
velocity and temperature laws and derive predicted line fluxes
as a function of the red giant mass loss rate, \mdot.  Our models 
generally match observations of the symbiotic stars EG And and 
AG Peg for \mdot~$\approx$ $10^{-8}~\msunyr$ to $10^{-7}~\msunyr$.
The optically thick cross-section of the red giant wind as viewed 
from the hot component is a crucial parameter in these models.
Winds with cross-sections of 2--3 red giant radii reproduce the 
observed fluxes, because the wind density is then high, $\sim$ 
$10^9$ \cm3.  Our models favor winds with acceleration regions
that either lie far from the red giant photosphere or extend 
for 2--3 red giant radii.

\vskip 4ex

{\it Subject headings:} binaries: symbiotic -- radiative transfer --
stars: emission-line -- stars: late-type

\eject

\vskip .6cm
\centerline{\sc {1.  Introduction}} 
\vskip 0.4cm
 
Symbiotic stars are long period, 1--100 yr, interacting binary systems 
consisting of a red giant, a hot companion, and a partially ionized 
emission nebula (Kenyon 1986).  Some material from the evolved giant, 
lost via a wind or tidal overflow, accretes onto the hot component.  
In many systems the hot component burns the accreted material and ionizes the 
red giant wind to produce a typical nebular emission spectrum.
The hot component also ionizes material close to the red giant and,
in some cases, the high density red giant atmosphere.  This illumination
produces many strong emission lines and a characteristic reflection
light curve (e.g., Kenyon, 1986; Kenyon \etal 1993; 
Miko{\l}ajewska \etal 1995; Nussbaumer, Schmutz, \& Vogel 1995). 

This paper continues our study of illuminated red giants in symbiotics.
In Proga \etal (1996; Paper I hereafter), we approximated a typical 
symbiotic with a hydrostatic red giant atmosphere illuminated by a 
point source.  This simple model yields a lower limit to illumination,
because the red giant atmosphere has a small scale height and intercepts
a small fraction of the high energy photons emitted by the hot component.
We calculated the atmospheric structure and emergent spectrum of the 
illuminated giant for a wide range of hot component effective temperatures 
and luminosities.  These models produce recognizably symbiotic spectra for
reasonable hot component temperatures, $\simgreat 10^5$ K, and luminosities,
$\simgreat 10^3 \lsun$.  Although our models also predicted electron densities 
and temperatures close to those estimated in many symbiotics, our predicted
emission line fluxes fall below observed fluxes by factors of 10--100.

In this paper, we consider the structure of illuminated red giant winds 
in symbiotic binaries.  Paper I's results show that a hydrostatic red giant 
is too small and does not intercept enough radiation from the hot component
to produce a bright illumination spectrum.  A red giant wind has a larger
cross-section than a hydrostatic giant and can thus intercept more high
energy photons from the hot component.  Paper I showed that an order of
magnitude increase in the cross-section should bring model fluxes close
to observations provided the density in the wind remains high,
$\sim 10^9 \cm3$.  Our goal is to calculate the one dimensional structure
of various wind models to test this hypothesis.

We describe the wind calculations and results 
in \S2.1--2.5 and compare our predictions with some observations
in \S2.6. We conclude with a brief discussion and summary
in \S3. 

\vskip .6cm
\centerline{ \sc {2. Models}}
\vskip 0.4cm

\vskip .6cm
\centerline{ 2.1. \it{ Background}}
\vskip .4cm

Most symbiotic binaries with red giant primaries have 
prominent permitted and intercombination emission lines 
superimposed on strong optical and ultraviolet continua. 
The intensity ratios of various intercombination lines -- such as
O~IV]$~\lambda1401$, S~IV]$~\lambda1406$, and N~III] $\lambda1750$ --
indicate high electron densities, $n_e \sim 10^9$ to $10^{10}~\cm3$, and 
large volume emission measures, $n_e^2V \sim 10^{59}$ to $10^{60}~\cm3$
(e.g., Kenyon 1986; Nussbaumer \& Stencel 1987; Kenyon \etal 1993). 
The high densities and narrow line profiles favor an emission
region close to the cool component for many species with low ionization
potentials, such as H~I, He~I, and O~III] (e.g., 
Nussbaumer \& Vogel 1990; Kenyon \etal 1993;  
Miko{\l}ajewska \etal 1995; Nussbaumer \etal 1995).
This emission region has a large size, $\rm R \sim$ 1--2 $\times~10^{13}$ cm,
compared to the scale height of a hydrostatic red giant atmosphere,
$z \sim 10^{11}$ cm. These emission lines thus form in a low velocity
red giant wind (e.g., Kenyon 1986; Nussbaumer \etal 1988; 
Kenyon \etal 1993; Munari 1993; Proga, Miko{\l}ajewska \& Kenyon 1994; 
Paper~I).

Other observations also indicate a substantial red giant wind in symbiotic 
stars.  For many systems, radio emission requires the hot component to
ionize a very large portion, $\rm R \simless$ 1--2 $\times~10^{15}$ cm,
of this wind (Seaquist, Taylor, \& Button 1984; Taylor \& Seaquist 1984).  
The inferred mass loss rates, $\mdot \sim 10^{-7}$ to 
$10^{-5} ~ \msunyr$, are appropriate for a red giant star or an evolved Mira 
variable (see also Whitelock 1987).  Analyses of the neutral portion of 
the wind yield similar results.  In particular, Vogel (1990, 1991) 
derived \mdot~$\approx$ 1.5 $\times ~ 10^{-8} ~ \msunyr$ and 
a wind terminal velocity of $v_{\infty} \approx$ 30 \kms~from 
observations of Rayleigh scattering in the extended atmosphere 
of the red giant in EG And.   Finally, the X-ray emission in some 
symbiotics appears to require colliding winds, in which high velocity
gas ejected from the hot component shocks a low velocity red giant wind 
with \mdot~$\approx$ 1--10 $\times ~ 10^{-7}$ \msunyr~(Willson \etal 1984;
Kwok \& Leahy 1984; Jordan, M\"urset, \& Werner 1994; 
Vogel \& Nussbaumer 1994; M\"urset, Jordan, \& Walder 1995).

To predict line and continuum emission from an illuminated red giant wind,
we first identify models that yield electron densities and emission 
measures appropriate for the ionized nebulae in symbiotic stars.
Although both $n_e$ and $n_e^2V$ are sensitive to many wind and hot
component parameters, we can place good constraints on the wind structure 
by deriving the radial distance of the ionization front from the 
center of the red giant, $r_{i}$, for a representative hot component.  
We set $n_e$ equal to the density at the ionization front and the
emitting volume equal to the product of the geometric cross-section 
and the density scale height, $l$: V $\approx \pi r_{i}^2 l$.  
For $n_e \approx$ $10^9$--$10^{10}$ \cm3, we require 
$r_{i} \sim 3~R_g$ and $l \sim A - r_i \simgreat R_g$, 
where $A$ is the orbital semimajor axis and
$R_g$ is the red giant radius (see also Paper I).

To estimate $r_{i}$, we follow Taylor \& Seaquist (1984; see also
Seaquist \etal 1984; Nussbaumer \& Vogel 1987) and balance 
ionizations and recombinations in the red giant wind along the 
line connecting both binary components.  
The appendix describes our calculation in more detail.
We first consider two simple wind models with parameterized 
velocity laws\footnote{We model winds with isothermal velocity
laws in \S2.5 to investigate the importance of thermal expansion.
These models predict emission line fluxes comparable to those
for other velocity laws and thus do not change the conclusions
derived from simple velocity laws in this section (Figures 9-10).}.
The first law is 
\begin{equation} 
v(r) = v_{\infty} \left(1-\frac{R_0}{r}\right)^\gamma,
\end{equation}
where $ v_{\infty}$ is the terminal velocity and
$R_0$ is a reference point at the base of the wind.  
This relation is commonly used in stellar wind studies
(e.g., Pauldrach, Puls, \& Kudritzki 1986; 
Kirsch \& Baade 1994; Harper \etal 1995) 
and reduces to a constant velocity wind when $\gamma$ = 0.
For the second velocity law, we adopt Vogel's (1991)
empirical relation,
\begin{equation}
{\frac{v(r)}{v_\infty}} = \left\{ \begin{array}{ll}
c_1 (r/R_g)^m               & {\rm for}~~\,~~  r/R_g~~\leq~~\hat{r}  \\
1 - exp[- c_2 (r/R_g -c_3)] & {\rm for}~~\,~~  r/R_g~~>~~\hat{r} \\
\end{array}  
 \right.
\end{equation}
where $m$ and $\hat{r}$ are the fitting parameters.
The parameters $c_1, c_2$, and $c_3$ allow both branches 
of the velocity to match smoothly at $r = \hat{r}R_g$
(i.e., $c_1={v(\hat{r} R_g)}/{v_\infty}/ \hat{r}^{m}$,
$c_2=m c_1 {\hat{r}^{m-1}}[{1-{v(\hat{r}R_g)}/{v_\infty} }]^{-1}$,
$c_3=\hat{r} + c_2^{-1}\ln[1-{v(\hat{r}R_g)}/{v_\infty}]) $ . 
This velocity law -- which has a smaller acceleration region and
an extended low velocity region compared to equation (1) -- fits 
eclipse observations of the red giant wind in the symbiotic EG And.
Finally, we adopt representative parameters for a symbiotic binary: 
$R_0 = R_g = 85~\RSUN$ for the red giant;
$T_h = 10^5$~K, $L_h = 600~\LSUN$ for the hot component;
$T_e = 3 \times 10^4$~K, the He abundance relative the H
abundance, $a(\rm He) = 0.1$, $v_\infty = 35$~\kms~for
the wind; and  $A = 3$~AU for the binary separation.

Our results show that the ionization front lies close to
the red giant photosphere for reasonable mass loss rates,
\mdot~$\simless~10^{-6}~\msunyr$ (Figure 1a).  For the 
$\gamma$-velocity law, the ionization front lies inside the
red giant photosphere when $\gamma$ = 0 and \mdot~$\simless~10^{-7}~\msunyr$
(solid line in Figure 1a; see also Taylor \& Seaquist 1984; 
Nussbaumer \& Vogel 1987).  The ionization front moves outside
the photosphere when $\gamma \ge$ 1 (dotted curve) and is at
$r_{i} \approx$ 2 $R_g$ when $\gamma \ge$ 3 (dashed curve).  
The variation of $r_{i}/R_g$ with \mdot~is less pronounced 
for Vogel's velocity law with $\hat{r} = 3.75$, $m=6$, and 
$v(\hat{r}R_g) = 0.3~v_\infty$ (dot-dashed curve).  
The ionization front lies at $r_{i} \ge$ 2 $R_g$ for 
\mdot~$\ge~10^{-8}~\msunyr$.  

In principle, both types of velocity laws yield ionized nebulae
with the physical conditions needed to match observations of
symbiotic stars.  Figure 1(b) shows the electron density at the
ionization front as a function of mass loss rate for a $\gamma$ = 0
wind (solid curve), a $\gamma$ = 3 wind (dot-dashed curve) and 
a wind with Vogel's velocity law (solid curve).  
The density at the ionization front exceeds $n_e \sim 3~\times~10^{9}$ 
\cm3~for all mass loss rates.  The average density in the ionized
wind is $<n_e>~\simgreat 10^9$ \cm3~for \mdot~$\ge 10^{-7}~\msunyr$
(Figure 1(c)).  These nebulae produce the needed $n_e^2 l \approx 10^{32}$ 
cm$^{-5}$ to $10^{33}$ cm$^{-5}$ for any velocity law with high enough 
mass loss rates, \mdot~$\ge~10^{-7}~\msunyr$ (Figure 1(d)).  If we
further require $r_{i} \ge$ 3 $R_g$ to yield $n_e^2 V \approx$
$10^{59}$ \cm3, we infer \mdot~$\approx~10^{-7}~\msunyr$ for Vogel's
velocity law and \mdot~$\approx~10^{-6}~\msunyr$ for $\gamma$ = 0
velocity laws.  These mass loss rates generally agree with estimates
derived from the radio nebulae of symbiotic stars 
(Seaquist, Krogulec, \& Taylor 1993).

We thus conclude that wind models with \mdot~$\simgreat~10^{-7}~\msunyr$
and either a Vogel-type velocity law or a velocity law with $\gamma$ = 3
can roughly match the observed densities and emission measures needed
for symbiotic nebulae.  Our simple calculation implies higher mass
loss rates for velocity laws with $\gamma \le$ 3.  We now consider 
detailed non-LTE photoionization calculations to test the ability of 
these models to match observed line fluxes in detail.  Non-LTE wind models 
are necessary due to the presence of an external radiation field 
that controls the 
emissivity of the ionized wind.  These models yield more accurate 
intensities for a wider range of atomic species than do LTE models 
(see also Paper I).

\vskip .6cm
\centerline{ 2.2. \it{ Non-LTE Photoionization Calculations}}
\vskip .4cm

In principle,  the ionization structure of an illuminated red giant
wind requires two dimensional calculations due to the large extent of 
the ionized region.  These calculations are straightforward if the
nebula is optically thin (e.g., Nussbaumer \& Vogel~1987), but
fully two-dimensional, non-LTE calculations are very complex and
require much computational time.  Here, we use a one-dimensional technique 
and include non-local processes in solving the radiative transfer problem. 
Our main goal is to quantify the main differences between spectra 
from illuminated hydrostatic red giant atmospheres and illuminated 
red giant winds.  A better understanding of these differences is
required before moving onto more complex calculations. 

To calculate the wind structure and spectrum, we use a 
modified version of Paper I's non-LTE photoionization code.
We assume a plane-parallel geometry for wind material 
illuminated by an external radiation field:
\begin{equation}
J_{\theta=0}^{in} (\nu,z)= \exp[-\tau_c(\nu,z)] 
\frac{L_h}{\sigma~T_h^4}\frac{1}{4\pi~(A-R_g-z)^2} 
B(\nu,T_h),
\end{equation} 
where $\tau(\nu,z)$ is the continuum optical depth at frequency
$\nu$ and height $z$
(compare equation A1 and A2 in Paper~I with equation 3 here).
We calculate the wind density profile assuming spherical 
symmetry (see below).
Our calculation divides the wind into 500 layers
with constant geometric thickness, $\Delta z$,
and iteratively solves for the emergent radiation 
field and the equilibrium temperature.
The density structure is fixed for wind models with a 
temperature-independent velocity, as in equations (1) and (2).
We consider models with an iterated density structure in \S2.5.
We apply a local escape probability method to solve the radiative
transfer equation in 500 logarithmically spaced bins covering
photon energies  0.01~eV~$\leq~h\nu~\leq$~1~keV with an 
energy resolution of log~$\delta (h\nu)=$~0.01.

To derive an equilibrium model for the illuminated wind, 
we calculate the structure from the top of the atmosphere,
$z_{top}$, to the point where 99.9\% or more of the incident 
flux with $\lambda \le 912$~\ang~has been absorbed, $z_{bottom}$.  
Each iteration consists of one or two steps,
depending on our assumption about the density structure.  
We first solve for the temperature, the level populations of 
each ionization state of twelve elements, and the radiation field
using energy balance and the local escape probability.
We then use the new temperature structure to integrate the 
continuity equation to derive a new density structure. 
This process repeats until the density and temperature converge to 
the 1\% level, which usually requires a few to 30 iterations.  
The final 1D model predicts the temperature and density structure 
with $z$ and the line and continuum emission.  
We simply add the red giant flux to the upward propagating flux 
from the illuminated wind,  because the wind optical depth is 
very low for this radiation.  Appendix A of Paper~I describes 
additional details of the model calculations.

Our ``basic'' wind model is similar to Paper~I's initial hydrostatic model.
We assume $T_g=3600$~K, $R_g= 85~\RSUN$, $M_g = 2.5~\MSUN$, and 
solar abundances (Anders \& Grevesse 1989) for the red giant and 
a binary separation, $A = 3$~AU, which are typical parameters for 
symbiotic stars (Kenyon 1986).  We adopt a constant mass loss rate,
$\MDOT = 10^{-6}$~\msunyr, and set $v_{\infty}=35$~\kms, $R_0 = R_g$,
and $\gamma = 3$ for the velocity law in equation (1).
Our atmosphere is initially isothermal with $z_{top}$ =
$3.14 ~ \times ~ 10^{13}$~cm and T = 3600~K. 
The mass loss rate and velocity law set the initial density,  
which remains fixed throughout the temperature iteration.  This 
approximation neglects pressure changes in the outflow due to 
the radial increase in temperature (see \S2.5) but is a reasonable 
first approximation to the actual situation.  

To calculate the total flux of the photoionized red giant wind,
we need to sum contributions from concentric annuli centered
on the substellar point of the red giant as viewed from the hot component
(see Figure 1 of Paper I).  We define $\phi$ as the angle subtended 
by each annulus and thus require models with 0 $\le \phi \le \pi$ for
a complete solution.  In this one-dimensional approach to the problem, 
we consider models with $\phi = 0$ and multiply the line and 
continuum emission by the geometric cross section of the red giant wind.  
The total spectrum of the model is then
\begin{equation}
F(\lambda) = \frac{L_h}{\sigma T^4_h}~B(\lambda,T_h) + 
4\pi \left\{R_g^2 B(\lambda,T_g) + r_{cs}^2 J^u(\lambda)\right\} ~,
\end{equation}
where $B(\lambda)$ is the photospheric emission from the 
hot and cool components, $J^u$ is the wind emission, and
$r_{cs}$ is the cross-sectional radius of the emission region
as viewed from the hot component.

The expression for the emitted flux, equation (4), differs somewhat from 
the analogous expression, equation (2) of Paper I, for hydrostatic models.
In Paper~I, we considered models where the red giant atmosphere is 
illuminated at a representative incident angle, $\theta_r$, for 
which the effective temperature of the illuminated annulus is equal 
to the effective temperature of the whole illuminated red giant hemisphere
(see Appendix~B in Paper~I).  This approximation is reasonable, because 
the scale height of the atmosphere is small (see Figure~1 of Paper~I).
We then derived a good approximation to the total flux from the 
illuminated hemisphere by multiplying the flux from this representative
annulus by the cross-section of the red giant photosphere.
These approximations fail in the present context due to the large 
geometrical extent of the wind.  First, the incident angle is a 
strong function of a distance from the red giant center.  
Unlike the hydrostatic case, the accuracy of our radiative transfer 
solutions for wind models decrease considerably as this incident angle
increases.  We thus only consider cases only along the line of centers
between the red giant and the hot component.  Our choice of the
appropriate geometric cross-section is also dictated by the large
scale height of a wind: a larger scale height implies a larger
cross-section (see Figure~1 of Paper~I).  Our photoionization
calculations in \S2.1 confirm that $r_{cs}~\sim~2-3~R_g$, which 
places the ionization front well above the red giant photosphere. 
We therefore adopt cross-sections derived from the ionization models 
shown in Figure~1 to scale our model fluxes.

Our assumption that $\phi=0$ (i.e., $\theta=0$ in Paper I, 
see Figure~1 and equation A2 in Paper~I) 
overestimates $z_{bottom}$, because the illumination
does not penetrate as far into the giant wind for $\phi \neq 0$. 
Consequently, the ionized wind, on average, is denser for 
$\phi=0$ than for $\phi\neq 0$.  Our models then underestimate
the predicted flux for regions close to the hot component and 
might overestimate the predicted flux for high density regions 
near the red giant. Our approximation that $r_{cs}$ is constant 
with $z$ has a similar effect on the predicted fluxes.
We estimate that the fluxes will change by a factor of $\simless 4$
with fully 2-D photoionization calculations.  
The difference between fully 2-D and our calculations  
is (i) the 1-D approximation assumed here (introducing an uncertainty
what we estimate is a factor of $\simless 2$) and (ii) 
the approximate method of
solving the radiative transfer problem (which also introduces a factor 
of $\sim 2$ uncertainty compared to a more realistic
non-LTE radiative transfer solution
(Avrett \& Loeser 1988,  see also Paper~I). Our results thus provide 
sensible first estimates for line fluxes from an illuminated wind. 
Although observations require 
$r_{cs} \approx 3 R_g$ as discussed above, we adopt $r_{cs} = R_g$ 
in this section to allow an easier comparison of wind models with
Paper I's hydrostatic models.  We consider more accurate estimates
for $r_{cs}$ when we compare our models with observations in \S2.6

As in our hydrostatic models, illumination divides the red giant wind
into a neutral lower atmosphere, a photoionized upper atmosphere, and 
a narrow transition region where ions recombine.  Figure~2 shows this 
structure for $T_h = 2 \times 10^5$~K and $L_h = 6.2 \times 10^4$~\LSUN.
Photoionization heats the upper atmosphere to
$\sim 6~\times~10^4$~K at large $z$ and maintains this high temperature
for 1.5 $\times~10^{13}$~cm $\simless z \simless$ 3 $\times~10^{13}$~cm. 
The temperature declines as the density increases with decreasing $z$,
because the ionization parameter decreases, leading to a smaller
heating per particle.  The temperature
has a small plateau at $\sim 3~\times~10^4$~K, drops rapidly to 
$\sim 2~\times~10^4$~K when He~III recombines to He~II, and then falls 
dramatically to its base level of 3600~K as He~II and H~II recombine.  
This structure qualitatively follows our hydrostatic models, although
the peak temperature in this wind model is roughly twice that of a 
hydrostatic model with identical $T_h$ and $L_h$ (see Figure~2 in Paper~I).

This illuminated wind produces more emission than the hydrostatic model.
A comparison of Figure~2 with Figure~2 in Paper~1 shows that the average
$n_e$ for an illuminated wind is lower than  an illuminated hydrostatic 
atmosphere by a factor of 10.  The depth of the ionized wind is also
several hundred times larger than the hydrostatic atmosphere:
$<n_e>_{wind} \approx 0.1 <n_e>_{HSE}$ and 
$l_{wind} \approx$ a few $ \times 10^2~l_{HSE}$.
This difference leads to a rough relation between the emission measure 
in both cases: $n_e^2l_{wind} \approx$ a few $n_e^2l_{HSE}$.
The volume emission measure for wind models is further 
increased by the ratio of geometric cross-sections: 
$n_e^2V_{wind} \approx$ a few $n_e^2 r_{cs}^2 l_{HSE}$
$\approx$ 10--30 $n_e^2 V_{HSE}$ for $r_{cs} \approx$ 2--3 $R_g$.
The He~II and He~III zones are also much larger due to the greater extent of 
the ionized wind, which absorbs more photons at a lower density (Figure~2d).  
Figure~3 -- which plots the total UV and optical flux of the model at a 
distance of 1 kpc -- shows the extra emission from the wind model quite 
clearly.  The reprocessed emission now contributes a significant portion 
of the UV and optical continuum flux.  Both the Balmer and Paschen continua
are roughly 5 times stronger than in a hydrostatic atmosphere.  The UV
and optical emission lines are also very intense.  The He~II $\lambda$4686
flux is  comparable to H$\beta$, as observed in many symbiotic stars
(Kenyon 1986).
The UV spectrum  includes strong He~II and N~V lines, which are very
weak in  hydrostatic models. In particular, some
high ionization UV lines, such as  Ne~VI($\lambda1009$) and 
Si~IV($\lambda1396$),  are  prominent even though they 
are practically invisible in a hydrostatic atmosphere. 
Generally, high ionization lines are more enhanced in wind models
compared to low ionization lines, because the highest emission 
measure material is closer to the hot component  where the 
UV radiation is stronger.

\vskip .6cm
\centerline{ 2.3. \it{ Model Grid}}
\vskip .4cm

We now consider a grid of wind models in the ($T_h$, $L_h$) plane.
The input hot component temperatures range from $0.2\times10^5$~K to
$2\times10^5$~K; the input luminosities span
$L_h \sim 6.2\times10^2~\LSUN$ to $6.2\times10^5~\LSUN$.
This grid repeats hot component parameters considered for 
the hydrostatic models in Paper~I and closely matches temperatures 
observed in most symbiotic stars.
The luminosity range of the model grid extends beyond typical hot
component luminosities but provides a good test of the code.
As in Paper~I, $\beta$ measures the strength of the illuminating 
radiation field relative to that of the giant at $\theta = 0^{\rm o}$:
\begin{equation}
\beta = \frac{R_g^2}{A^2}\frac{L_h}{L_g} ~.
\end{equation}
We also adopt $r_{cs} \approx R_g$ as in \S2.2.

Figure~4 compares blackbody flux curves for an unilluminated binary
with model spectra for $T_h$ = 0.5, 1, and 2 $\times~10^5$~K and
$\rm log~\beta$ = $-$2 to 1. Hydrostatic models produce strong emission 
lines only for $T_h \simgreat  10^5$~K.   
For a  low temperature hot component, 
wind models now  have a strong H$\alpha$ emission line and several strong UV 
emission lines (Figure~4a).  
These models also have a weak Balmer emission jump and some He~I emission
lines not visible in hydrostatic models.  The richness of the 
emission spectrum increases with $T_h$ (Figure~4b and 4c).  Prominent
He~I and He~II optical lines appear at $T_h = 10^5$~K; these lines and the 
Balmer continuum strengthen considerably at $T_h = 2 ~ \times ~ 10^5$~K
(Figure~4c).  The UV line emission is also very sensitive to $T_h$ and
dominates the UV spectrum at $T_h = 2 ~ \times ~ 10^5$~K.  At all temperatures,
the contrast between the UV lines and UV continuum decreases with
increasing $L_h$.  Many lines saturate at high $L_h$; the strength of
the Balmer continuum and intrinsic hot component flux continue to 
increase with $L_h$.

Wind models also produce larger changes in the  optical and 
infrared magnitudes 
than corresponding hydrostatic models.  Figure~5 shows the magnitude
differences -- $\Delta \rm U$, $\Delta \rm B$, $\Delta \rm V$, and 
$\Delta \rm K$ -- between illuminated and non-illuminated hemispheres 
of the red giant.  Practically all wind models have a measurable 
magnitude difference, $\Delta \rm U \simgreat$ 0.1 mag, between 
the two hemispheres.  The most luminous models have large differences, 
$\Delta \rm U \simgreat$ 0.2 mag, comparable to those observed in 
some symbiotic stars (Kenyon 1986; Munari 1993).  
The largest $L_h$ models also have significant infrared 
differences, $\Delta \rm K \approx$ 0.4 mag, {\it not} observed 
in symbiotics.  This behavior indicates that our highest luminosity 
models are too luminous for typical symbiotic stars.

Predicted fluxes for most lines are a factor of $\sim$ 10 stronger 
at the highest temperatures and luminosities in our grid (Table 1;
see also Figures 7--9 of Paper I).
This factor increases towards lower $T_h$ or $L_h$ due to the smaller
geometric dilution factor.  The extended red giant atmosphere absorbs 
more photons at lower density than in corresponding hydrostatic models 
where the dilution factor is 4--9 times larger.  The emission line 
equivalent widths of wind models are thus much closer to those observed 
in many symbiotic stars (Figure~6; Table~2).  The UV line equivalent 
widths compare well with observations for high temperature models with
$T_h \simgreat 10^5$~K.  The H$\beta$ and He~II $\lambda$4686 lines now reach 
typical observed values of EW(H$\beta$) $\approx$ EW(He II $\lambda$4686) 
$\approx$ 50--100~\ang~for several high luminosity models.  Lower luminosity
models, which are a better match to the luminosities of most symbiotics,
manage EW(H$\beta$) $\approx$ 1--10~\ang.  This EW is too small to explain 
the line flux of most real systems.  However, predicted equivalent widths 
increase to levels observed in many symbiotics for $r_{cs} \approx$ 
2--3 $R_g$, as discussed in \S2.6 below.

The variations of line ratios with ($T_h$, $L_h$) generally follow those 
predicted for hydrostatic models described in Paper~I 
(compare Figure~10 of Paper~I with Figure~7 here).  
The H~I line ratios shown in Figure~7 vary considerably 
because optical depth effects
are very important for the high electron densities, 
$n_e \sim 10^{9} - 10^{10}~\rm \cm3$, that characterize the red giant
wind
(Figure~8; see also Paper~I and Drake \& Ulrich 1980, for example).  
The I(H$\alpha$)/I(H$\beta$) 
intensity ratio far exceeds the case B ratio at low luminosities and 
accidentally is close to the case B value at high luminosities.  
The I(He II $\lambda$1640)/I(He II $\lambda$4686) ratio lies close 
to 10 for most combinations of ($T_h$, $L_h$) but decreases by as
much as a factor of 1.5 at high $T_h$ (Figure~7; upper right panel).  
These line ratios are poor 
reddening diagnostics unless ($T_h$, $L_h$) is well-known (Paper I).

In contrast, wind models provide better matches to observations of
He~I line ratios (Figure~7; lower panels).  
The optical He~I lines are good diagnostics of the 
physical conditions in symbiotics, because the line ratios are
sensitive to the optical depth and the electron density in the ionized
nebula (Almog \& Netzer 1989; Schmid 1989; Proga \etal 1994). 
Our wind models generally yield higher ratios for 
I($\lambda$6678)/I($\lambda$5876) and I($\lambda$7065)/I($\lambda$5876) 
compared to hydrostatic models.  
In particular, predictions for I($\lambda$7065)/I($\lambda$5876) match
observed ratios -- log I($\lambda$7065)/I($\lambda$5876) $\approx$ $-$0.3
to 0.0 -- for hot components with $T_h \simgreat$ 30,000~K and $\beta$
= $-$2 to 1.  Hydrostatic models fit the full range of observed values
only for unrealistically high $\beta$.
Although some wind models with very luminous hot components now reach 
the log I($\lambda$6678)/I($\lambda$5876) $\approx -0.1$ observed in 
some symbiotics, most models with reasonable $\beta$ fail to match
observations.  Paper I suggested that a lower density, more extended 
model atmosphere with a better treatment of the He$^+$ recombination zone 
near the red giant photosphere would produce intensity ratios closer
to those observed.  Our results confirm the first half of this 
prediction; models with better atomic and molecular opacities are
needed to test the second half.

Figure~8 shows mean parameters for global physical conditions
within the photoionized wind.
The average temperature of the H~II region is significantly higher
than for the hydrostatic case, because a large portion of the wind
lies very close to the hot component. The temperature of 
the recombined He~II 
region in both cases is similar, because this region always lies 
close to the red giant photosphere.  The geometric dilution factor 
also varies considerably for the H~II region and changes very little 
for the He~II region.

The mean electron density of the H~II region,
$n_e \sim 10^9-10^{10}$~\cm3, is roughly a factor 
of 10 smaller in wind models than in hydrostatic models.
Two factors cause this behavior.
The lowest density in  hydrostatic models is much
higher than the lowest density in wind models 
(see Figure~2 and Paper~I's Figure~2, for example).
Wind models also have more material at large 
distances from the red giant than  hydrostatic models.
Consequently, low density regions contribute more to 
$<n_e>$ in wind models than in hydrostatic models. 
The mean electron density also does not increase as rapidly with $L_h$.
We keep the density structure fixed in wind models, so
more luminous hot components simply ionize larger, higher density
portions of the wind compared to lower luminosity hot components.
In contrast, our density iteration in hydrostatic models allows the 
density in the H~II region to increase with increasing $L_h$ to maintain
hydrostatic equilibrium.  

\vskip .6cm
\centerline{ 2.4. \it{ Wind models with Vogel's velocity law.}}
\vskip .4cm

To illustrate the sensitivity of our results to the adopted
wind velocity law, we now consider illuminated winds with 
Vogel's (1991) empirical relation (equation [2]).
Figure 9 compares Vogel's velocity law with several $\gamma$
velocity laws for $v_{\infty}$ = 35 \kms.  The $\gamma$ and Vogel
velocity laws all converge to $v_{\infty}$ at $r \gg 10^{13}$ cm, 
but the velocity near the hot component, $r \approx$ 3--5 
$\times 10^{13}$ cm, is largest for Vogel's velocity law.  
The velocity decreases as $\gamma$ increases.
The mass density near the hot component is thus lowest for Vogel's 
velocity law and increases with $\gamma$.  Although this difference 
is not large, $\sim$ 50\% at $r \sim 4 \times 10^{13}$ cm, high
energy photons for {\it low luminosity} hot components penetrate
farther into a wind with Vogel's velocity law than into a wind with
a $\gamma$ velocity law.  Despite the extra path length of the 
ionized region, the lower density in most of the ionized region  
reduces $n_e^2l$ for Vogel's velocity law. We thus expect {\it smaller}
emission line fluxes from a Vogel-type velocity law compared to 
$\gamma$ laws for low luminosity hot components.  

This behavior reverses as the hot component luminosity increases.
For $r \simless 2 \times 10^{13}$ cm, the wind velocity is smaller
for the Vogel velocity law than for $\gamma$ laws (Figure 9).
The wind density thus increases considerably with depth for the Vogel
law compared to the $\gamma$ laws.  The difference in the electron 
densities at $r \sim$ 1--2 $\times 10^{13}$ cm is a factor of 5--10 
and leads to a considerable difference in the path length through 
the ionized region.  In general, the density at the ionization front 
and the mean density in the ionized nebula, $<n_e>$, are larger
for Vogel-type velocity laws than for $\gamma$ laws, although the 
path length through the nebula is smaller (see Figure 1a).  The
emission line flux scales as $n_e^2 l$, so we expect Vogel wind 
models to have {\it larger} emission line fluxes than $\gamma$ 
wind models for {\it high luminosity} hot components.

Figure 10 compares predicted fluxes
in H$\beta$, He~I $\lambda$5876, 
O~III] $\lambda$1664, C~IV $\lambda$1550, He~II $\lambda$4686, and 
N~V $\lambda$1240  
for different wind models (Tables~1 and 3).
We  adopt $T_h$ = $10^5$~K, $r_{cs} = R_g$ in every case.  
Our chosen 
emission lines span a wide range of ionization states that probe 
a variety of density and temperature regimes within the 
outflowing wind.
Figure 10 confirms our general expectations from a simple analysis
of the velocity laws in Figure 9. 
The $\gamma$ = 3 models should predict 
larger fluxes for the highest ionization lines such as He~II and N~V 
when $\beta$ is small, because the small number of He$^+$ and N$^{+3}$
ionizing photons do not penetrate into a significant fraction of the 
red giant wind.  The larger numbers of H~I and He~I ionizing photons 
penetrate closer to the red giant photosphere; H~I and He~I line
fluxes for Vogel and $\gamma$ = 3 velocity laws are then comparable.
For both sets of emission lines, line fluxes from $\gamma$ models
lag behind Vogel-type models as the hot component luminosity increases.  
The Vogel-type models produce 
comparable N~V fluxes for $\beta \simgreat -2$ and rival $\gamma$ = 3
models at higher mass loss rates for $\beta \simgreat$ 0.  The
difference between the Vogel and $\gamma$ velocity laws is most 
apparent for lower ionization lines, such as H~I and He~I, that form 
throughout the wind.  The predictions for H$\beta$ and $\lambda$5876
with a Vogel-type velocity law are comparable to $\gamma$ laws at much 
higher mass loss rates for $\beta \simgreat$ 0 due to the abrupt
transition in the velocity at $r \sim$ 1--2 $\times 10^{13}$ cm.

Figure 10 demonstrates that the mass loss rate inferred from
optical and ultraviolet emission line fluxes clearly depends on
the adopted velocity law.  The uncertainty in \mdot~for luminous 
hot components is large, $\sim$ a factor of 10, when we use  
lower ionization features such as He~I, O~III], and C~IV as wind
diagnostics.  The higher ionization lines, He~II and N~V, are
better diagnostics, because these lines form closer to the hot
component where the red giant wind is closer to terminal velocity.
The local density and line fluxes then depend primarily on \mdot~for
fixed $L_h$ and $T_h$.  Any wind from the hot component, however,
will modify the density law considerably and may make it difficult
to constrain \mdot~from emission lines.

\vskip .6cm
\centerline{  2.5. \it{``Isothermal'' wind}}
\vskip .4cm

In \S2.3 and \S2.4 we assume that the velocity and density
laws of the red giant wind are not functions of the wind temperature. 
However, the wind density and velocity profiles change
with wind temperature due to  thermal expansion. 
Stronger illumination increases the pressure term in the 
wind momentum equation and thus changes the wind dynamics.
An analogous  situation occurs in hydrostatic models, where illumination
regulates the density structure through the temperature and ionization 
(Paper~I).
To take this process into account, we adopt the steady state, 
isothermal wind model used for the solar wind (e.g., Parker 1958; 
Mihalas 1978).
Figure~2 illustrates that $T_e$ 
is almost constant for a large fraction of the H~II region
and is very close to the $<T_e>$ of this region.
We therefore approximate the temperature structure 
by a constant temperature, $T_0$,  in our calculations of the wind velocity.
The velocity of the isothermal wind is:
\begin{equation} 
\left(\frac{v}{a}\right)^2 - {\rm ln} \left(\frac{v}{a}\right)^2 = 
4~{\rm ln} \frac{r}{r_c} + 4~\frac{r}{r_c} - 3
\end{equation}
where $a$ is the isothermal sound speed
\begin{equation}
 a=\left(\frac{2kT_0}{\mu m_H}\right)^{1/2}
\end{equation}
and $r_c$ is the sonic point
\begin{equation}
 r_c=\frac{GM_g \mu m_H}{4kT_0}.
\end{equation}
We assume $\mu=1$ in this simple, isothermal treatment of thermal expansion.
As in \S2.2, we calculate the radial density distribution from
the continuity equation once we specify $\mdot$.
For a typical $T_0=30,000$~K and the
red giant parameters as in \S2.2, $a=22~\kms$ and $r_c= 5.7 R_g$. 

Figure~9 compares the velocity laws calculated from equations 
(1), (2), and (6--8). 
Solid lines represent  velocities at different temperatures 
using equation (6); the velocity at large radii increases with $T_0$.
Dashed lines represent velocities for different $\gamma$ 
using eq. (1); $\gamma$ decreases with increasing velocity.
The isothermal velocities are clearly very sensitive to $T_0$ 
and increase without limit as $r$ increases.  This behavior 
contrasts with other velocity laws that converge asymptotically
to the terminal velocity.  Although radio data for symbiotic stars
preclude a velocity that continues to increase at large radii
(Seaquist \etal 1984), the isothermal laws for $T_0=2-4\times10^4$~K 
provide good lower and upper bounds to the $\gamma$ and Vogel 
velocities for $r \simless 4 \times 10^{13}$~cm.  An isothermal wind
thus yields reasonable density structures for the region between the
hot and cool components of a typical symbiotic system.

To predict the spectra of an illuminated isothermal wind, we follow
\S2.2 but modify the initial conditions and computational procedure.
For the initial conditions, we adopt $z_{top}=3.14 \times 10^{13}$~cm 
and an initial $T_0=3\times10^4$~K. After completing the calculations 
of the equilibrium ionization and temperature structure, we adopt the
average electron temperature of the photoionized wind, $<T_e>$, 
as the new $T_0$ for equations (7--8) (see Figure 8), derive a new 
velocity law from equation (6), and calculate a new density profile from 
the continuity equation.  Our iteration scheme otherwise proceeds as 
in Paper I's hydrostatic models (see Appendix A of Paper~I). 
Thus, the wind temperature and ionization structure are from 
the non-LTE photoionization calculations despite the isothermal 
wind approximation used to calculate the wind velocity and density.

Figure 11 compares predictions for isothermal wind models (crosses)
with models for the Vogel velocity law (triangles) and a $\gamma$ = 3
velocity law (diamonds) for \mdot~= $10^{-7} \msunyr$.  
Isothermal models predict lower emission fluxes than the other models.
For $\beta \simgreat -1$, the isothermal wind temperature is
$T_0 \simgreat$ 30,000 K and the wind velocity is larger than wind
velocities for either the Vogel velocity law or any of the $\gamma$ 
velocity laws (Figure 9).  The density at the ionization front is
thus lower in isothermal models, which allows photons to penetrate
more deeply into the wind compared to $\gamma$ law models.  
This lower mean density in the wind leads to the lower line fluxes 
for isothermal models.  The isothermal wind temperature remains high
as the luminosity of the hot component decreases, so the mean wind
density -- and hence the predicted line fluxes -- also remain low 
compared to models with other velocity laws.  

In contrast to models with Vogel's velocity law, isothermal models 
predict {\it higher} emission line fluxes as $\beta$ and $T_h$
decrease.  The isothermal wind temperature is then relatively small,
$T_0 \simless$ 30,000 K, which leads to higher mean densities in
the wind compared to the $\gamma$-law models.  This comparison shows 
once again that an accurate knowledge of the wind velocity law is
needed to estimate \mdot~from observations.  In addition to the
obvious changes in the density structure, isothermal winds predict
different $r_{cs}$ at fixed \mdot~compared to models with $\gamma$
or Vogel-type velocity laws.

Our simple, isothermal wind models include thermal expansion but 
neglect the kinetic energy of the wind in the energy equation.
This approximation is appropriate for high densities typical
for our models because the time scale for radiative cooling,
$t_{rad}$, is much shorter than the dynamic time scale, $t_{dyn}$. 
The minimum density is $\sim 10^8~\rm cm^{-3}$ for models with
$\MDOT=10^{-7}~\msunyr$, For $T_e= 30,000$~K, the thermal energy,
$\sim 10^{-3}~\rm erg$, and the cooling rate, 
$2.8\times10^{-7}~\rm erg~s^{-1}$, yield $t_{rad} \approx 3600$~s.
The dynamical time is $t_{dyn}= 2.5 \rm AU/ 35 ~\rm km~s^{-1} = 10^7$ s. 
Thus the temperature is fixed by radiative processes.
 
\vskip .6cm
\centerline{ 2.6. \it{Comparison between different models and observations}}
\vskip .4cm

Having described several types of wind models, we now consider
comparisons with data for two symbiotic systems, AG Peg and EG And.
EG And contains a low luminosity hot component, $L_h \approx 16~\lsun$,
that produces prominent UV emission lines but weak optical emission lines
(Kenyon 1986; M\"urset \etal 1991; Vogel, Nussbaumer, \& Monier 1992).  
Vogel (1991) used eclipse light curves of the hot component to derive
the mass loss rate and velocity law of the cool component; Vogel \etal
(1992) measured the radius of the cool component from these data.
The symbiotic nova AG Peg has a higher luminosity hot component,
$L_h \approx 600~\lsun$, that excites a substantial ionized nebula.
The system is also one of the more luminous radio symbiotic stars
(Kenny, Taylor,  \& Seaquist 1991 and references therein).  Both systems
are good examples of a red giant wind illuminated by the hot component.
Kenyon \etal (1993) show that many emission lines form in the outer
atmosphere of the red giant in AG Peg.  Munari (1993) describes
evidence that a heated red giant atmosphere produces H$\alpha$ and 
several UV emission lines in EG And.

Our comparison of models with observations has two primary uncertainties, 
the velocity law of the wind and the cross-sectional area of the giant.  
Figure~1a and our non-LTE models show that the ionization front lies
outside the red giant photosphere for $\gamma \simgreat~2$ and Vogel's
velocity law when $\MDOT~\simgreat~10^{-8}$~\msunyr~and $\phi$ = 0\degree.
We derive similar results for $\phi >$ 0.  The radius of the ionization
front increases from $r_i \approx 2.7 R_g$ at $\phi$ = 0\degree~to
$r_i \approx 3.2 R_g$ at $\phi$ = 25\degree~for parameters appropriate
to AG Peg.  The electron density at the ionization front decreases from 
$n_e(r_i) \sim 7 \times 10^9$ \cm3~to $\sim 2\times 10^9~\rm cm^{-3}$ as 
$\phi$ increases from 0\degree~to 25\degree.  For $\phi \simgreat 25\degree$, 
both $r_i$ and $n_e(r_i)$ change much faster with $\phi$, because the 
density does not increase monotonically with distance from the hot 
component (see also the lower panel of Figure~2 in Taylor \& Seaquist 1984).
These results indicate $r_{cs} \approx 3 R_g$ as a reasonable approximation
for the cross-sectional radius of the ionization front in AG Peg.
We similarly estimate $r_{cs} \approx 2.5 R_g$ for EG And, a system with 
a lower luminosity hot component and a lower red giant mass loss rate.
These adopted cross-sections increase our predicted fluxes by factors
of 5--10 compared to models where the ionization front is coincident 
with the red giant photosphere.

The wind velocity law introduces additional uncertainty into comparisons
with observations, as described in \S2.2--2.5.  For this paper, we adopt 
Vogel's velocity law.  This relation is consistent with observations of
at least one red giant wind, EG And, and provides a good lower limit
on the red giant mass loss rate needed to match observations (Figures
10--11).  As we show below, the mass loss rates we derive for EG And
and AG Peg are close to those estimated from radio observations.  
In addition,  Vogel's velocity law predicts an extended low velocity
region that is lacking in $\gamma$-type velocity laws.  Extended
low velocity regions are  characteristic of some theoretical 
red giant wind models (e.g., van Buren, Dgani, Noriega-Crespo 1994 
and references therein).
 
Figure~12 compares line fluxes for wind models and observations of AG~Peg.
We use the data from Paper I (Kenyon \etal 1993; see Altamore \& Cassatella 
1997 for a re-discussion of the UV evolution of this system).
We adopt $T_h = 10^5$~K and $L_h=620~\LSUN$ as best hot component 
parameters for AG Peg and use red giant and wind parameters from \S2.2 for 
Vogel's velocity law with $\MDOT~=~10^{-8},~10^{-7}$, and $10^{-6}~\msunyr$.  
Open triangles show predictions for $r_{cs} = R_g$; filled triangles 
show predictions for $r_{cs} = 3 R_g$.  The line fluxes increase with 
\mdot~for each set of models.  The predicted line fluxes vary roughly
linearly with \mdot~at high \mdot, because the emission measure of the
ionized wind increases linearly with \mdot~(see Figure 1d).  At lower
mass loss rates, the line fluxes are limited by the amount of material
in the {\it static} red giant atmosphere.  This material provides a
lower limit to the emission measure -- for a particular hot component
luminosity -- so the line fluxes vary little for \mdot~$\simless$
$10^{-8}~\msunyr$ (Figure 1; see also Paper I).

Considering the many uncertainties, our models provide a reasonable 
match to the observations.  Radio observations indicate \mdot~$\sim$
$10^{-7}~\msunyr$ for the cool component (Kenny \etal 1991; Kenyon 
\etal 1993).  Our models fall below all of the observations for
$r_{cs} \approx R_g$ and generally bracket the data for 
$r_{cs} \approx 3 R_g$.  In particular, we predict higher than
observed fluxes for the moderate ionization lines, O~III] and C~IV,
and predict lower than observed fluxes for low and high ionization lines.
A good match to the O~III] and C~IV fluxes depends on our assumption
of $r_{cs} \approx 3 R_g$ and our adoption of solar abundances for
the red giant photosphere.  Reducing the O/N and C/N abundances to
values more typical of field red giants -- $\sim$ 6--10 times less
than solar (Smith \& Lambert 1985, 1986) -- would bring our predictions
closer to observations.  Reducing our adopted cross-section to 
$r_{cs} \sim 2.5~R_g$ would also yield a better match to the data.
In view of the uncertainties in the abundances (Nussbaumer \etal 1988;
Whitelock \& Munari 1992) and the cross-section, we did not attempt
to find best-fit values for these parameters.

Better fits to observations of other lines require a better wind model.
Although a modest abundance increase can yield higher N~V fluxes,
abundance variations are not responsible for low H~I, He~I, and He~II
fluxes (see also Paper I).  As noted in Paper I, we expect some extra 
He~I emission from the recombination region close to the red giant 
photosphere.  Our calculations do not include H$^-$ and molecular 
opacities and thus underestimate He~I radiation from this region.
We need a higher density in the ionized wind or a more luminous hot 
component or both to match data for the H~I and He~II emission lines.
Our good match to O~III] and C~IV emission probably precludes a
significant change to our adopted $L_h$.  Colliding wind models
can yield higher densities close to the red giant (see below)
and are a possible source of extra H~I and He~II emission
(Nussbaumer \& Vogel 1989; Nussbaumer \& Walder 1993).  

Figure~13 compares model line fluxes with observations of EG~And. 
The H~I and He~II lines are now H$\alpha$ and He~II~$\lambda$1640. 
As in Figure 12, we adopt Vogel's velocity law with parameters in
\S2.2 and $\MDOT~=~10^{-8},~10^{-7}$, and $10^{-6}~\msunyr$.  
We also adopt $L_h \sim 16~\LSUN$, $T_h \sim~7.5~\times~10^4$~K,
$L_g \sim 950~\LSUN$,  $T_g \sim 3700$~K, $R_g \sim 75~\RSUN$,
$A = 1.5$~AU, a 400 pc distance, and $E_{B-V} =  0.05$
(Vogel \etal 1992).  Open triangles show predictions 
for $r_{cs} = R_g$; filled triangles show predictions for $r_{cs} = 2.5 R_g$.
As in AG Peg, the model fluxes reach a minimum level at the lowest
mass loss rates due to the emission measure set by the hydrostatic 
atmosphere.  This limit is most noticeable in lower ionization lines 
that form throughout the ionized wind at higher mass loss rates,
such as H~I and He~I.  The observations include {\it IUE} data at 
orbital phase 0.465 (Vogel \etal 1992) re-reduced for this study and
ground-based optical data from KPNO (see Kenyon \etal 1993~for
a description of KPNO optical data).  
We adopt I(He~I$\lambda$5876)/I(H$\alpha$)~$\sim~0.01-0.1$ as typical
for S-type symbiotics in the absence of a real detection of this line.

As with AG Peg, our models bracket observations for $r_{cs} > R_g$.
For the best estimate of the mass loss rate, 
$\MDOT \approx 10^{-8}$~\msunyr (Vogel 1991), we can match 
observed fluxes only for C~IV and N~V if $r_{cs} \approx R_g$.
Observations of other lines exceed predictions by factors of 3--10.
The model predictions exceed all of the observed fluxes for 
$r_{cs} \sim 3 R_g$ when $\MDOT \approx 10^{-8}$~\msunyr.
Model and observed fluxes agree if we reduce both $r_{cs}$ and 
the abundances of CNO elements.  For such low mass loss rates,
illumination greatly modifies the wind velocity and density through
thermal expansion, and probably reduces $r_{cs}$ as outlined in \S2.5.
A modest reduction in the cross-section to $r_{cs} \approx 2 R_g$
would bring the predicted H~I, He~I, and He~II fluxes into reasonable 
agreement with the observations.  Predicted fluxes for the CNO elements
then exceed observed fluxes by factors of 3--5, which we interpret as
a low metal abundance.  The high negative radial velocity, $\sim$
$-$100 \kms, and high galactic latitude, $b \approx -22\degree$,
indicate that EG And is a member of the halo population and likely
to have lower metal abundances than adopted in our models
(Schmid \& Nussbaumer 1993).  Other
symbiotics, such as AG Dra, have similarly low metal abundances
(Smith, Cunha, Jorissen, \& Boffin 1996).

\vskip .6cm
\centerline{\sc {3. Discussion}}
\vskip .4cm

Our non-LTE models support Paper I's conclusion that an illuminated
red giant wind can produce optical and ultraviolet emission line 
fluxes similar to those observed in most symbiotic stars.  Winds 
with \mdot~$\approx$ $10^{-7}~\msunyr$ generally match observations 
for AG Peg; lower mass loss rates can match observations of EG And.  
In both systems, the large ionized volume of a low velocity wind is 
necessary to explain data for H~I and He~II.  The cross-section of 
the red giant wind as viewed from the hot component is a crucial 
factor for low ionization emission lines, such as He~I, C~IV, and O~III].  
Winds with large cross-sections, $r_{cs} \approx$ 2--3 $R_g$, 
reproduce the observed fluxes, because the wind density at 2--3 $R_g$ 
is $\sim$ $10^9$ \cm3~(see also Paper I).  High ionization lines, 
such as N~V, require wind material close to the hot component, 
where the dilution factor is small, or a higher abundance to match
the observed fluxes.

We commented in \S2.2 that our scaling of model fluxes by a 
constant cross-section probably underestimates predicted fluxes
of lines formed close to the hot component and overestimates
fluxes of lines formed close to the red giant photosphere.
Observed fluxes for both EG And and AG Peg support this comment:
high ionization lines are somewhat stronger than low ionization
lines in real symbiotics compared to the model results.
Although a better treatment of the two-dimensional nature of 
the problem is required to address this issue in more detail,
our model results compare reasonably well with observations 
in both cases.

Wind models provide some constraints on the wind velocity law if
the mass loss rate is well-known.  We can generally rule out 
$\gamma$ velocity laws with $\gamma \approx$ 0--1 if 
$\mdot \sim 10^{-7}$ \msunyr.  Radio observations preclude the 
higher mass loss rates that allow $r_{cs} \approx$ 2--3 $R_g$ 
for $\gamma \approx$ 0--1.  Both $\gamma$ = 3 and Vogel
velocity laws yield the densities and emission measures derived
for symbiotic stars.  Most wind properties for these two classes 
of models are similar and thus do not discriminate between models.
In particular, both models produce high densities, 
$\simgreat 10^9~\cm3$, at distances of 2--3 $R_g$ from the 
center of the red giant.  Nevertheless, Vogel's velocity law 
predicts a higher cross-section than $\gamma$ = 3 models and 
thus produces better fits to observations of EG And and AG Peg.

Our one dimensional wind model works best for less active symbiotic 
stars like EG And.  An empirical wind velocity law from eclipse
observations -- together with well-known system parameters from
UV and optical studies -- provide good constraints on the system
geometry.  The weak or nonexistent wind from the hot component
eliminates the complications of colliding winds in systems like
AG Peg, where the system geometry is also less certain.  

The He~I line ratios are the major failure of our wind models.
As we noted in Paper I, He~I emission is very sensitive to the
electron density and optical depth through the wind and red
giant atmosphere (see also Proga \etal 1994).  We suspect that
radiation from the coolest parts of the recombination region
contributes much He~I emission.  Although we treat this portion
of the atmosphere better than in Paper I, we still do not exhaust 
hot component photons for $\lambda > 912 \rm ~\AA$ and do not include 
the H$^-$ and molecular opacities needed to model well the red 
giant photosphere.  We thus underestimate the He~I fluxes of
all lines and also underestimate several important line ratios.
Schwank, Schmutz, \& Nussbaumer (1997) support these conclusions.
Their illuminated red giant winds produce H~I emission lines in a 
dense, narrow hydrogen recombination zone with $T_e \simless 10^4$ K.
Accurate treatment of this emission requires good molecular 
opacities and high radial resolution not included in our models.

Other deficiencies in our models require a multi-dimensional
treatment.  The typical, complicated symbiotic system is not
well-approximated by a one dimensional wind, even when we 
include the wind cross-section as an additional model parameter.
Axisymmetric solutions with accurate two-dimensional ionization
structures should provide better estimates of line fluxes for
low ionization features -- e.g., He~I, C~IV, and O~III] -- in
systems like AG Peg.  These types of models could also predict 
line {\it profiles} for comparison with observations.  Although 
many low ionization lines have narrow, symmetric profiles, some
lines have extra absorption and emission features that could
diagnose properties of an illuminated wind (e.g., Munari 1993;
Schwank \etal 1997).

The high ionization features He~II and N~V, as well as H~I emission, 
may require more complicated wind geometries.  Colliding winds can
provide an extra source of ionization and also raise the density 
in the outflowing wind (e.g., Wallerstein \etal 1984; Kwok 1988).
Colliding winds in EG And and AG Peg probably contribute little 
additional luminosity to the line emission, because the kinetic 
energy in the winds is small compared to the hot component 
luminosities ($l_{wind} \sim 10~\lsun$ in AG Peg and 
$l_{wind} \ll 1~\lsun$ in EG And).  Higher wind densities 
can raise the optical depth through the wind -- and 
hence increase $r_{cs}$ -- and increase collisional excitation.
Both of these processes are important for H~I and high ionization 
emission lines, where our models currently underpredict line fluxes. 
In principle, the line profiles and fluxes are good diagnostics 
of these types of models, but these models also introduce additional 
free parameters (e.g., Nussbaumer \& Vogel 1989; Nussbaumer \& Walder 1993).  
Our results indicate that simple changes to the wind velocity law 
can also produce large changes in model line fluxes.  
Further exploration of colliding winds {\it and} single wind models 
with different velocity laws may clarify which lines are most 
sensitive to which model parameter (see Contini 1997).

The emitted spectrum of the hot component is another uncertainty
of our model.   An input blackbody spectrum overestimates the
numbers of H-ionizing and He-ionizing photons compared to static 
white dwarf atmospheres, which have strong absorption edges at 
912~\ang, 554~\ang, and 228~\ang.  We suspect, however, that the 
hot component in AG Peg has a strong wind and more closely resembles 
a Wolf-Rayet star (see Kenyon \etal 1993; Nussbaumer \& Walder 1993).
By analogy with the models in  Crowther, Smith, \& Hillier (1995), 
we expect small H~I absorption edges and possibly some He absorption 
edges.  The depths of these edges depend on the exact temperature, 
wind structure, and chemical composition of the atmosphere. 
Absorption edges in the spectrum 
of the hot component will clearly change the ionization structure 
in the red giant wind and the emergent emission fluxes. In particular, 
strong He~II edges reduce abundances of highly ionized species in the
wind (e.g., He~II and N~V) and enhance low ionization line fluxes 
(e.g., He~I).  The large uncertainty in the appropriate effective
temperatures for symbiotic hot components precludes quantifying the
importance of absorption edges in real systems, so we have ignored 
these features in this study.

Despite these deficiencies, our models demonstrate that illumination 
modifies the structure of the red giant wind in most symbiotic
stars.  We identify two main portions of the wind: a neutral wind
where material acceleration begins and an ionized wind where the
material reaches terminal velocity.  For systems like AG Peg, 
radiation from the hot component raises the temperature in 
the ionized wind to $T \sim$ 30--60 $\times~10^3$ K in regions 
beyond the sonic point at $r \simgreat$ 2--3 $R_g$ (see Figure 8).  
The winds are nearly isothermal at these radii.  Isothermal wind 
models with $T \sim$ 30--40 $\times~10^3$ K predict fluxes close 
to those observed in symbiotic stars.  These models indicate that 
thermal expansion alone can accelerate the wind to the observed 
velocities.  Indeed, thermal expansion may dominate the dynamics 
of the ionized wind.

Our models suggest that illumination may {\it not} modify the
red giant mass loss rate.  High energy photons from the hot
component penetrate close to the red giant photosphere only
when \mdot~is low or $L_h$ is very high.  Neither of these
situations is very typical of symbiotic stars.  In those cases 
where high energy photons from the hot component reach the red giant 
wind acceleration zone, the increase in the temperature and the 
decrease in density of the illuminated atmosphere probably 
reduces the mass loss rate of winds driven by acoustic or 
magnetic waves as we discussed in Paper~I.  Illumination will also
reduce the mass loss rates of winds driven by radiation pressure on
dust and abundant molecules, because radiation from the hot
component reduces the abundances of dust and molecules in the
red giant atmosphere.  We expect increases in the mass loss rate 
only when the illumination flux increases on a time scale short 
compared to the cooling time scale of the red giant, $\sim$ 1 hr.
The red giant atmosphere then expands 
and the density gradient flattens (Paper I).  Once the illumination
flux becomes steady, i.e. after $t_{dyn}$, the red giant mass loss 
rate probably settles to a value smaller than or equal to the mass 
loss rate established prior to the rise in the illumination flux.

Our models also suggest that illumination can modify the rate 
of accretion onto the hot component.  Illumination from the
hot component heats the red giant wind.  The heating rate is
proportional to the hot component luminosity; the wind velocity 
thus increases with increasing $L_h$.  In contrast, the mass accretion
rate onto the hot component, \mdot$_h$, is a strong {\it inverse}
function of the wind velocity in wind-accreting systems
(e.g., Tutukov \& Yungel'son 1982; Livio \& Warner 1984).
For example, \mdot$_h$ is roughly proportional to $v_{W}^{-4}$ if
the red giant wind velocity is constant (Livio \& Warner 1984;
their equation 2).  Detailed multi-dimensional hydrodynamical 
calculations show that the accreting star primarily captures
downstream wind material (see Ruffert 1996 and Benensohn \etal 1997).
Although we do not consider this portion of the wind in our
calculations, we expect that illumination will increase the 
downstream velocity of the wind and reduce the accretion rate 
as discussed above.

An accretion rate modulated by illumination has important consequences 
for variability of the hot component on several different time scales.  
For example, the time-variation of \mdot$_h$ is chaotic in many 
hydrodynamical simulations of wind accretion, with factor of two 
changes on time scales of hours to days (see Benensohn \etal 1997 
and references therein).  If $L_h$ keeps pace with \mdot$_h$,
illumination will probably enhance \mdot$_h$~variations due to the
negative feedback between $L_h$ and $v_W$.  This feedback could lead
to observable changes in emission line fluxes on short time scales.
A strong negative feedback between $L_h$ and \mdot$_h$~could also lead to 
large-scale instabilities in the accretion flow.  Bell, Lin, \& Ruden (1991) 
identified a radiative-feedback mechanism in accretion disks and 
showed that this mechanism could produce instabilities and irregular 
outbursts in the accretion disks of pre-main sequence stars.  Although 
the disks in wind-fed symbiotic systems are different from the disks 
in pre-main sequence stars, outbursts in the two types of systems share
common features.  As in dwarf novae, mass transfer rates into the disk 
would need to lie in a narrow range of unstable accretion rates to give
rise to the instability (Lin \& Papaloizou 1996).  

Illumination may play an important role in the long-term evolution 
of the hot component even if there are no instabilities in the accretion
flow.  Thermonuclear runaways are the most promising outburst mechanism
for the symbiotic novae, where 3--6 mag eruptions last for many decades
(Kenyon 1986).  If illumination changes the time-averaged mass accretion 
rate of the hot component significantly, then the time scale for a system
to initiate a symbiotic nova eruption will also change.  An increase
in the eruption time scale {\it reduces} the number of symbiotic nova
eruptions during the lifetime of the red giant donor and thus reduce
the amount of nuclear-processed material added to the hot component
during the red giant's lifetime.  Symbiotics become less likely progenitors
for type Ia supernovae if these reductions are large (see Yungel'son
\etal 1995; Iben \& Tutukov 1996).

Finally, our results demonstrate that illumination probably plays 
an important role in the dynamics of other interacting binaries.
The $\zeta$ Aur and VV Cep systems -- binaries with K-M supergiant
primaries and O-B main sequence secondaries -- have many features
in common with symbiotics, including accretion from low velocity stellar
winds (e.g., Eaton 1994; Bauer \etal 1991).  Illumination probably
leads to variations in the accretion flow close to the hot components
and might be responsible for the short term variations in emission line
fluxes and absorption line equivalent widths observed in systems like
VV Cep (Stencel, Potter, \& Bauer 1993).  Massive X-ray binaries also
show rapid variations in the spin period (Reig 1997).  These changes 
are often associated with changes in \mdot~(see Benensohn \etal 1997 
and references therein); our results indicate that illumination may
regulate the spin period by changing \mdot~in wind-accreting systems.

\vskip 6ex

Part of this work has been included in a doctoral dissertation presented
at Nicolaus Copernicus Astronomical Center in Warsaw, Poland.
We thank 
E. Avrett, J. E. Drew, R. Loeser, J. Miko{\l}ajewska, 
D. Sasselov, and H. Uitenbroek for helpful advice and comments.
We also thank an anonymous referee for comments that helped us 
clarify our presentation.
This project was partially supported by KBN Research Grants
No. 2 P03D 015 09 and 2 P304 007 06, National Aeronautics 
and Space Administration Grant NAG5-1709, the Smithsonian
Institution's Predoctoral Fellowship Program, and PPARC grant.

\eject
\Appendix{A}

The location of the photoionized front
in the red giant wind  can be estimated from 
the recombination--ionization balance in the H~II
region (e.g., Seaquist \etal 1984; Taylor and Seaquist 1984; Nussbaumer
\& Vogel 1987).  
Here we will repeat the  reasoning 
and formalism from these papers for a case
where the wind velocity is a smooth function of a distance from the giant.
We also use 
the same notation, except for the angular variable
and the separation which we call $\phi$  and $A$ instead of $\theta$  
and $p$, respectively.

We assume that the wind consists of hydrogen and helium;
in the H~II region, hydrogen is fully ionized and helium is singly ionized.
We use the on-the-spot approximation for the diffuse Lyman continuum photons. 
The recombination-ionization balance 
for the infinitesimal solid angle, $\Delta\phi$ in the direction
$\phi$ is
\begin{equation}
L_{\rm H} \frac{\Delta\phi}{4 \pi} = \Delta\phi \int_{0}^{s_\phi}
n(s)n_e(s)\alpha_B({\rm H},T_e) s^2 ds,
\end{equation}
where $s_{\phi}$ is the distance of the hydrogen recombination zone
for the angle $\phi$, $L_{\rm H}$ is the rate of 
the hot component photons with $\lambda \leq 912$~\ang, and $\alpha_\beta$
is the total hydrogen recombination coefficient for a case B.
We assume  a spherically symmetric stellar wind with a constant mass loss rate, 
$\dot{M}$. From the continuity equation, 
${\dot M} = 4\pi r^2 n_{\rm H}(r) \mu m_{\rm H} v(r)$, 
the radial density law is
\begin{equation} 
n_{\rm H}(r) = \frac {\dot M}{ 4\pi~r^2~\mu~m_{\rm H}~v(r)},
\end{equation}
where $\mu$ is the mean molecular weight, $m_{\rm H}$ is the mass of 
the hydrogen atom, and $n_{\rm H}$ is the hydrogen number density.
The electron density, $n_e$ is
\begin{equation}
n_e(r)=(1+a({\rm He})) n_{\rm H}(r),
\end{equation}
where a({\rm He}) is the helium abundance relative to hydrogen.
Introducing a new spatial variable $u=s/A$, and assuming an isothermal
H~II region, the equation (1) is
\begin{equation}
X^{{\rm H^+}} = v_{\infty}^2 \int_{0}^{u_\phi} \frac{u^2}
{\left((u^2 + 1 - 2 u \cos \phi) v(s,\phi)\right)^2}~du,
\end{equation}
where 
\begin{equation}
X^{{\rm H^+}} = \frac{4 \pi \mu^2 {m_{\rm H}}^2}{\alpha_B({\rm H},T_e)
(1 + a({\rm He}))} A L_H 
\left(\frac{v_\infty}{\MDOT}\right)^2.
\end{equation}

\vfill
\eject


\eject

\centerline{\bf Figure Captions}

\vskip 4ex
\noindent
Figure~1 -- Physical characteristics of a simple model wind
as a function of $\MDOT$ along a direction {$\phi = 0$}
for representative parameters of a symbiotic binary (see \S2.1).
Results are for the  $\gamma~=~0$ (solid curves), 
1 (dotted curves), 3 (dashed curves), and 
Vogel's velocity law (dot-dashed curves).
For clarity, not all curves are plotted in each panel.
(a) Top left panel: location of the ionization front in a giant wind.
Two horizontal long dashed lines show  range of Roche lobe radii
for most symbiotic stars.
For reference, the binary separation is $A/R_g \approx 8$.
(b) Top right panel: electron density at the ionization front.
(c) Bottom left panel: mean electron density in the ionized wind. 
(d) Bottom right panel: emission measure in the ionized wind.

\vskip 4ex
\noindent
Figure~2 -- Model wind structure for $T_h=2\times10^5$~K,
$\beta = 1$, $\gamma=3$ and $\MDOT=10^{-7}$~\msunyr. 
(a) Top left panel:
$T_e$ radial profile.
(b)~Top right panel:
hydrogen and electron number density profiles
(solid and dashed lines).
(c)~Bottom left panel: fractional population of the ground levels
of H~I and H~II (solid and dashed lines).
(d)~Bottom right panel: fractional population of
the ground levels
of  He~I, He~II, and He~III  (solid, dotted, and dashed lines).

\vskip 4ex
\noindent
Figure~3 -- Total UV and optical flux of the model wind (solid line)
for a distance of 1~kpc, an orbital inclination of $\rm 90^o$, and
orbital phase of 0.0 (when the hot component lies in front of the giant).
For comparison, the dotted and dashed lines show the flux of 
a model illuminated hydrostatic atmosphere (dotted curve) and
two blackbody flux curves without illumination (dashed curve) 
for the same parameters.

\vskip 4ex
\noindent
Figure~4 -- Blackbody flux curves for a binary without illumination
(thin lines) and model wind spectra (thick lines) for different $\beta$
and $T_h$ = 0.5, 1.0, and $2.0\times10^5$~K, panel a, b, and c,
respectively.  We assume a distance of 1~kpc as in Figure~3.

\vskip 4ex
\noindent
Figure~5 -- Differences in broadband magnitudes between a red giant 
with and without an illuminated wind as a function of $T_h$.
The four thick curves indicate differences for non-LTE model parameters
in Figure~4 (log~$\beta$ = $-$2, $-$1, 0, and 1).  The thin dotted horizontal
line plots magnitude differences for an LTE model.

\vskip 4ex
\noindent
Figure~6 -- Equivalent widths  for H$\beta$, He~I $\lambda$5876, 
O~III] $\lambda$1664, C~IV $\lambda$1550, He~II $\lambda$4686, and 
N~V $\lambda$1240  
as
functions of $T_h$ for log~$\beta$ = $-$2 (solid curves), $-$1 (dotted curves),
0 (dashed curves), and 1 (dot-dashed curves).

\vskip 4ex
\noindent
Figure~7 -- H and He emission line flux ratios as functions of $T_h$
for log~$\beta$ = $-$2 (solid curves), $-$1 (dotted curves),
0 (dashed curves), and 1 (dot-dashed curves).

\vskip 4ex
\noindent
Figure~8 -- Mean parameters describing the global physical
conditions within the photoionized red giant wind as functions
of $T_h$ for log~$\beta$ = $-$2 (solid curves), $-$1 (dotted curves),
0 (dashed curves), and 1 (dot-dashed curves).
Panels in the left column plot mean values for the complete
wind; panels in the right column plot
values for the He~II region.

\vskip 4ex
\noindent
Figure~9 -- Comparison of velocities from the $\gamma$, Vogel's 
and  isothermal laws. 
Solid lines show isothermal law velocities at three temperatures, $T_0 = 2, 
4$, and $6 \times 10^4$~K; $T_0$ increases with increasing velocity. 
Dashed lines show
$\gamma$ law velocities for $\gamma = 1, 2,$ and 3; $\gamma$ decreases
with increasing velocity. Dot-dashed line shows for Vogel's
law velocity.
Two dashed vertical lines mark the region of our
computations ($ 0.~\leq~z~\leq~z_{top}$).

\vskip 4ex
\noindent
Figure~10 -- Comparison of model line fluxes from the wind models
for $\gamma=3$, $\MDOT=10^{-6}$\msunyr~(Figure~6) and $\MDOT=10^{-7}$\msunyr
with the wind model for Vogel's law and $\MDOT=10^{-7}$\msunyr.
Squares and diamonds show the line fluxes of the wind model for $\gamma=3$, 
$\MDOT=10^{-6}$ and $10^{-7}$~\msunyr, respectively.
Triangles show the line fluxes of the wind model for 
the Vogel's law, $\MDOT=10^{-7}$~\msunyr.
All models are for $T_h = 10^5$~K,
a distance of 1~kpc, and log~$\beta$ = $-$3, $-$2, $-$1, 0, and 1,
except the models for $\gamma=3$ and $\MDOT=10^{-7}$ models 
where log~$\beta$ = $-$3, $-$2, $-$1, and 0.
At each ion, $\beta$ increases with increasing line flux.

\vskip 4ex
\noindent
Figure~11 -- Comparison of model line fluxes from  the wind models
for $\gamma=3$, Vogel's, and isothermal velocity laws 
for $\MDOT=10^{-7}$\msunyr. 
Diamonds and triangles show the line fluxes of the wind model for $\gamma=3$, 
and Vogel's law, respectively.
Crosses show the line fluxes of the wind model for 
the isothermal law.
All models are for $T_h = 10^5$~K,
a distance of 1~kpc, and log~$\beta$ = $-$2, $-$1, 0, and 1,
except the models for $\gamma=3$ and $\MDOT=10^{-7}$ models 
where log~$\beta$ = $-$2, $-$1, and 0.
At each ion, $\beta$ increases with increasing line flux.

\vskip 4ex
\noindent
Figure~12 -- Comparison of model line fluxes from the wind models
for Vogel's law,  and two cross sections with
observations for AG~Peg. 
Open and filled  triangles show the line fluxes of the
wind model for the cross section $R_g^2$ (open triangles) and $3^2 R_g^2$
(filled triangles) for $T_h = 10^5$~K,
a distance of 1~kpc, log~$\beta$ = $-$2, and $\MDOT=10^{-6}, 10^{-7}$,
and $10^{-8}$\msunyr.
At each ion, $\MDOT$ increases with increasing line flux.
The stars connected by the dashed line indicate observations 
at d = 1~kpc corrected for reddening, $E(B-V)=0.1$~mag.

\vskip 4ex
\noindent
Figure~13 -- Comparison of model line fluxes from the wind models
for Vogel's law,  and two cross sections with
observations for EG~And. 
Open and filled  triangles show the line fluxes of the
wind model for the cross section $R_g^2$ (open triangles) and $2.5^2 R_g^2$
(filled triangles) for $T_h = 10^5$~K,
a distance of 1~kpc, log~$\beta$ = $-$3, and $\MDOT=10^{-6}, 10^{-7}$,
and $10^{-8}$\msunyr.
At each ion, $\MDOT$ increases with increasing line flux.
The crosses connected by the dashed line indicate observations 
at d = 1~kpc corrected for reddening, $E(B-V)=0.05$~mag.
For the He~I line crosses show possible
range of the fluxes, i.e., 0.1 - 0.01$\times$F(H$\alpha$).

\eject

\begin{center}
{Table 1. Logarithm of model emission line $\rm fluxes^{a}$
for $\gamma=3$ and $\MDOT=10^{-6}~\msunyr$.} 
\footnotesize
\vspace{2mm}
{
\begin{tabular}{c  c c c c c c c c c c c c c} 
\\ \hline \hline
$L_h$ &  $\beta$  & $T_h$ & H$\alpha$ & H$\beta$ & P$\alpha$ & He~I & He~I & 
He~I & He~I & He~I & He~I & He~II & He~II \\
$(\LSUN$) &     & ($10^5$~K) & & & &$\lambda$3889 &$\lambda$4471 &$\lambda$5876 
&$\lambda$6678 &$\lambda$7065 &$\lambda$10830 &$\lambda$1640 & $\lambda$4686 \\ 
\hline 
$6.2\times10^2$ & 0.01 &  0.2 & -11.0 & -11.6 & -12.0 & -12.9 & -13.4 & -12.9 & 
-13.5 & -13.5 & -12.3 & -18.2 & -19.2 \\
$6.2\times10^2$ & 0.01 &  0.5 & -10.4 & -11.2 & -11.4 & -12.0 & -12.1 & -11.4 & 
-12.2 & -11.5 & -10.5 & -12.1 & -13.1 \\
$6.2\times10^2$ & 0.01 &  1.0 & -10.4 & -11.3 & -11.3 & -12.0 & -12.1 & -11.4 & 
-12.2 & -11.5 & -10.5 & -10.8 & -11.8 \\
$6.2\times10^2$ & 0.01 &  2.0 & -10.6 & -11.5 & -11.5 & -12.4 & -12.7 & -12.2 & 
-13.0 & -12.5 & -11.1 & -10.4 & -11.5 \\
 & & & & & & & & & & & & & \\ 
$6.2\times10^3$ & 0.1  &  0.2 & -10.3 & -11.1 & -11.3 & -12.3 & -12.7 & -12.2 & 
-12.8 & -12.9 & -11.9 & -17.2 & -18.2 \\
$6.2\times10^3$ & 0.1  &  0.5 &  -9.7 & -10.6 & -10.6 & -11.2 & -11.6 & -10.6 & 
-11.5 & -10.6 & -10.0 & -11.3 & -12.4 \\
$6.2\times10^3$ & 0.1  &  1.0 &  -9.7 & -10.6 & -10.6 & -11.2 & -11.6 & -10.6 & 
-11.5 & -10.6 &  -9.9 & -10.1 & -11.1 \\
$6.2\times10^3$ & 0.1  &  2.0 &  -9.8 & -10.9 & -10.8 & -11.7 & -11.9 & -11.4 & 
-12.2 & -11.5 & -10.4 &  -9.7 & -10.8 \\
 & & & & & & & & & & & & & \\ 
$6.2\times10^4$ & 1    &  0.2 &  -9.5 & -10.2 & -10.4 & -11.6 & -12.1 & -11.6 & 
-12.2 & -12.3 & -11.3 & -16.0 & -17.0 \\
$6.2\times10^4$ & 1    &  0.5 &  -9.0 &  -9.6 & -10.0 & -10.0 & -10.7 &  -9.9 & 
-10.6 &  -9.9 &  -9.6 & -10.7 & -11.7 \\
$6.2\times10^4$ & 1    &  1.0 &  -9.0 &  -9.6 & -10.0 &  -9.8 & -10.6 &  -9.8 & 
-10.5 &  -9.8 &  -9.5 &  -9.4 & -10.4 \\
$6.2\times10^4$ & 1    &  2.0 &  -9.2 &  -9.8 & -10.1 & -10.4 & -11.1 & -10.2 & 
-11.0 & -10.2 &  -9.6 &  -8.9 &  -9.9 \\
 & & & & & & & & & & & & & \\ 
$6.2\times10^5$ & 10   &  0.2 &  -8.6 &  -8.8 &  -9.6 & -10.7 & -11.1 & -10.6 & 
-11.2 & -11.2 & -10.2 & -14.5 & -15.5 \\
$6.2\times10^5$ & 10   &  0.5 &  -8.5 &  -8.8 &  -9.6 &  -9.2 &  -9.7 &  -9.3 &  
-9.5 &  -9.4 &  -9.1 & -10.0 & -11.0 \\
$6.2\times10^5$ & 10   &  1.0 &  -8.5 &  -8.9 &  -9.6 &  -9.1 &  -9.6 &  -9.2 &  
-9.3 &  -9.3 &  -9.1 &  -8.5 &  -9.4 \\
$6.2\times10^5$ & 10   &  2.0 &  -8.7 &  -9.1 &  -9.7 &  -9.3 &  -9.9 &  -9.4 &  
-9.8 &  -9.5 &  -9.2 &  -8.0 &  -8.9 \\
\hline
\end{tabular}
}
\end{center} 
\eject 

\normalsize
\centerline{ Table 1. Continued }
\footnotesize
\vspace{-6mm}
\begin{center}
\begin{tabular}{c  c c c c c c c c c c} \\ \hline \hline
$L_h$ &  $\beta$  & $T_h$ & C~II & C~II] & C~III] &  C~IV & N~II] & N~III] & 
N~IV] & N~V \\
($\LSUN$) &  & ($10^5$~K)   &$\lambda$1335 &$\lambda$2325 &$\lambda$1908 
&$\lambda$1550 &$\lambda$2141 & $\lambda$1750 & $\lambda$1486 & $\lambda$1240 \\ 
\hline
$6.2\times10^2$ & 0.01 &  0.2 &  -11.8 & -10.9 & -11.5 &       & -12.3 & -12.8 & 
      &       \\ 
$6.2\times10^2$ & 0.01 &  0.5 &  -12.2 & -11.7 &  -9.9 & -10.2 & -13.4 & -10.9 & 
-11.4 & -14.4 \\ 
$6.2\times10^2$ & 0.01 &  1.0 &  -12.1 & -11.8 & -10.3 &  -9.3 & -13.8 & -11.4 & 
-10.5 & -10.3 \\ 
$6.2\times10^2$ & 0.01 &  2.0 &  -12.2 & -11.9 & -10.8 &  -9.7 & -13.9 & -12.0 & 
-11.2 & -10.3 \\ 
 & & & & & & & & & &  \\
$6.2\times10^3$ & 0.1  &  0.2 &  -11.2 & -10.3 & -10.7 & -16.4 & -11.6 & -11.9 & 
-17.4 &       \\ 
$6.2\times10^3$ & 0.1  &  0.5 &  -11.5 & -11.3 &  -9.3 &  -9.4 & -13.0 & -10.1 & 
-10.7 & -12.3 \\ 
$6.2\times10^3$ & 0.1  &  1.0 &  -11.6 & -11.5 &  -9.9 &  -8.7 & -13.8 & -10.8 & 
 -9.7 &  -9.6 \\ 
$6.2\times10^3$ & 0.1  &  2.0 &  -11.9 & -11.7 & -10.5 &  -9.0 & -14.0 & -11.5 & 
-10.5 &  -9.8 \\ 
 & & & & & & & & & &  \\
$6.2\times10^4$ & 1    &  0.2 &  -10.4 &  -9.6 & -10.1 & -14.9 & -10.9 & -11.1 & 
-16.0 &       \\ 
$6.2\times10^4$ & 1    &  0.5 &  -10.7 & -10.8 &  -8.7 &  -8.8 & -12.6 &  -9.1 & 
 -9.8 & -10.9 \\ 
$6.2\times10^4$ & 1    &  1.0 &  -10.7 & -11.0 &  -9.3 &  -8.1 & -13.4 &  -9.9 & 
 -8.8 &  -8.9 \\ 
$6.2\times10^4$ & 1    &  2.0 &  -11.3 & -11.4 & -10.2 &  -8.3 & -14.0 & -10.8 & 
 -9.6 &  -9.1 \\ 
 & & & & & & & & & &  \\
$6.2\times10^5$ & 10   &  0.2 &   -9.5 &  -8.9 &  -9.5 & -13.5 & -10.0 & -10.3 & 
-14.5 &       \\ 
$6.2\times10^5$ & 10   &  0.5 &   -9.8 & -10.3 &  -8.2 &  -8.2 & -12.0 &  -8.4 & 
 -8.8 & -10.0 \\ 
$6.2\times10^5$ & 10   &  1.0 &   -9.9 & -10.7 &  -8.8 &  -7.5 & -13.0 &  -9.0 & 
 -8.0 &  -8.1 \\ 
$6.2\times10^5$ & 10   &  2.0 &  -10.5 & -11.3 &  -9.6 &  -7.7 & -13.9 &  -9.9 & 
 -8.6 &  -8.4 \\ 
\hline
\end{tabular}
\end{center}


\eject
\normalsize
\centerline{ Table 1. Continued }
\footnotesize
\vspace{-10mm}
\begin{center}
\begin{tabular}{c  c c c c c c c c c c c} \\ \hline \hline
$L_h$ &  $\beta$  & $T_h$ & O~III] &  O~IV] & O~V] & O~V] & O~VI & Mg II & 
Si~II] & Si~III] & Si~IV \\
($\LSUN$) &     & ($10^5$~K) & $\lambda$1664 & $\lambda$1403 & $\lambda$1218 & 
$\lambda$1371& $\lambda$1034 & $\lambda$2800& $\lambda$2335& $\lambda$1892& 
$\lambda$1397 \\ \hline
$6.2\times10^2$ & 0.01 &  0.2 & -13.8 &       &       &       &       & -12.8 & 
-13.9 & -11.2 & -14.7 \\
$6.2\times10^2$ & 0.01 &  0.5 & -10.5 & -12.6 & -14.6 & -21.1 &       & -11.8 & 
-13.4 & -10.7 & -10.6 \\
$6.2\times10^2$ & 0.01 &  1.0 & -10.4 & -10.5 & -10.0 & -14.3 & -10.8 & -11.2 & 
-13.2 & -11.5 & -10.9 \\
$6.2\times10^2$ & 0.01 &  2.0 & -11.1 & -10.8 &  -9.8 & -13.8 &  -9.2 & -11.4 & 
-13.2 & -11.6 & -10.9 \\
 & & & & & & & & & & \\ 
$6.2\times10^3$ & 0.1  &  0.2 & -12.8 &       &       &       &       & -12.4 & 
-13.6 & -10.6 & -13.2 \\
$6.2\times10^3$ & 0.1  &  0.5 &  -9.7 & -11.2 & -12.4 & -17.6 &       & -11.3 & 
-13.3 & -10.2 &  -9.8 \\
$6.2\times10^3$ & 0.1  &  1.0 &  -9.5 &  -9.9 &  -9.3 & -13.4 &  -9.8 & -10.6 & 
-13.7 & -11.1 & -10.2 \\
$6.2\times10^3$ & 0.1  &  2.0 & -10.3 & -10.2 &  -9.0 & -13.0 &  -8.6 & -10.8 & 
-13.5 & -11.3 & -10.4 \\
 & & & & & & & & & & \\ 
$6.2\times10^4$ & 1    &  0.2 & -11.9 & -17.9 &       &       &       & -11.4 & 
-13.0 &  -9.9 & -11.9 \\
$6.2\times10^4$ & 1    &  0.5 &  -8.9 & -10.3 & -10.9 & -15.6 & -14.5 & -10.6 & 
-13.2 &  -9.4 &  -9.0 \\
$6.2\times10^4$ & 1    &  1.0 &  -8.6 &  -9.1 &  -8.6 & -12.6 &  -9.1 &  -9.8 & 
-14.1 & -10.5 &  -9.4 \\
$6.2\times10^4$ & 1    &  2.0 &  -9.2 &  -9.6 &  -8.2 & -11.9 &  -7.9 & -10.0 & 
-14.1 & -10.9 &  -9.7 \\
 & & & & & & & & & & \\ 
$6.2\times10^5$ & 10   &  0.2 & -11.2 & -16.8 &       &       &       &  -9.9 & 
-11.5 &  -9.2 & -10.9 \\
$6.2\times10^5$ & 10   &  0.5 &  -8.2 &  -9.6 & -10.1 & -14.5 & -13.0 &  -9.9 & 
-12.7 &  -8.8 &  -8.3 \\
$6.2\times10^5$ & 10   &  1.0 &  -7.8 &  -8.3 &  -7.9 & -11.4 &  -8.5 &  -9.2 & 
-13.7 & -10.1 &  -8.5 \\
$6.2\times10^5$ & 10   &  2.0 &  -8.3 &  -8.9 &  -7.5 & -10.4 &  -7.2 &  -9.1 & 
-14.7 & -10.4 &  -8.9 \\
\hline
\end{tabular}
\end{center}

\noindent
{Note.--Blank spaces mark fluxes practically equal to zero.}\\

\noindent
$\rm ^a$ {In units of erg~$\rm  cm^{-2}~s^{-1}$.} 

\eject
\normalsize
\begin{center}
{ Table 2. Logarithm of model emission line 
equivalent widths $\rm (EWs)^a$ for $\gamma=3$ and $\MDOT=10^{-6}~\msunyr$.} 
\footnotesize
{
\begin{tabular}{c  c c c c c c c c c c c c c} 
\\ \hline \hline
$L_h$ &  $\beta$  & $T_h$ & H$\alpha$ & H$\beta$ & P$\alpha$ & He~I & He~I & 
He~I & He~I & He~I & He~I & He~II & He~II \\
$(\LSUN$) &     & ($10^5$~K) & & & &$\lambda$3889 &$\lambda$4471 &$\lambda$5876 
&$\lambda$6678 &$\lambda$7065 &$\lambda$10830 &$\lambda$1640 
& $\lambda$4686 \\ \hline
$6.2\times10^2$ & 0.01 &  0.2 &   0.5 &   0.0 &   0.0 &  -1.3 &  -1.7 &  -1.3 &  
-2.0 &  -2.0 &  -0.7 &  -7.2 &  -7.6 \\
$6.2\times10^2$ & 0.01 &  0.5 &   1.2 &   0.6 &   0.7 &   0.1 &  -0.1 &   0.3 &  
-0.6 &   0.1 &   1.1 &  -0.7 &  -1.2 \\
$6.2\times10^2$ & 0.01 &  1.0 &   1.2 &   0.6 &   0.7 &   0.3 &  -0.1 &   0.3 &  
-0.7 &   0.1 &   1.1 &   1.3 &   0.1 \\
$6.2\times10^2$ & 0.01 &  2.0 &   1.0 &   0.3 &   0.5 &  -0.1 &  -0.7 &  -0.6 &  
-1.4 &  -0.9 &   0.5 &   2.3 &   0.5 \\
 & & & & & & & & & & & & & \\ 
$6.2\times10^3$ & 0.1  &  0.2 &   0.9 &  -0.1 &   0.7 &  -1.5 &  -1.8 &  -1.1 &  
-1.6 &  -1.7 &  -0.3 &  -7.2 &  -7.3 \\
$6.2\times10^3$ & 0.1  &  0.5 &   1.8 &   1.0 &   1.4 &   0.3 &   0.0 &   0.9 &  
 0.1 &   0.9 &   1.6 &  -0.9 &  -0.7 \\
$6.2\times10^3$ & 0.1  &  0.0 &   1.9 &   1.2 &   1.5 &   0.9 &   0.3 &   1.0 &  
 0.1 &   1.0 &   1.7 &   1.0 &   0.7 \\
$6.2\times10^3$ & 0.1  &  0.0 &   1.7 &   0.9 &   1.3 &   0.5 &   0.0 &   0.3 &  
-0.6 &   0.0 &   1.2 &   2.0 &   1.1 \\
 & & & & & & & & & & & & & \\ 
$6.2\times10^4$ & 1    &  0.2 &   0.9 &  -0.2 &   1.3 &  -1.9 &  -2.1 &  -1.3 &  
-1.7 &  -1.8 &  -0.2 &  -7.0 &  -7.1 \\
$6.2\times10^4$ & 1    &  0.5 &   2.2 &   1.4 &   2.0 &   0.7 &   0.1 &   1.3 &  
 0.7 &   1.3 &   2.0 &  -1.3 &  -0.8 \\
$6.2\times10^4$ & 1    &  0.0 &   2.4 &   1.9 &   2.1 &   1.5 &   0.9 &   1.7 &  
 0.9 &   1.6 &   2.1 &   0.7 &   1.1 \\
$6.2\times10^4$ & 1    &  0.0 &   2.4 &   1.9 &   2.0 &   1.4 &   0.7 &   1.4 &  
 0.5 &   1.3 &   2.0 &   1.9 &   1.9 \\
 & & & & & & & & & & & & & \\ 
$6.2\times10^5$ & 10   &  0.2 &   0.8 &   0.3 &   1.4 &  -1.9 &  -2.1 &  -1.3 &  
-1.7 &  -1.6 &   0.0 &  -6.5 &  -6.5 \\
$6.2\times10^5$ & 10   &  0.5 &   1.9 &   1.2 &   2.2 &   0.5 &   0.1 &   0.9 &  
 0.9 &   1.1 &   1.9 &  -1.6 &  -1.1 \\
$6.2\times10^5$ & 10   &  0.0 &   2.4 &   1.8 &   2.3 &   1.3 &   1.0 &   1.6 &  
 1.6 &   1.7 &   2.2 &   0.5 &   1.2 \\
$6.2\times10^5$ & 10   &  0.0 &   2.5 &   2.0 &   2.3 &   1.7 &   1.2 &   1.7 &  
 1.4 &   1.7 &   2.3 &   1.8 &   2.2 \\
\hline
\end{tabular}
}
\end{center}

\eject
\normalsize
\centerline{ Table 2. Continued }
\footnotesize
\vspace{-6mm}
\begin{center}
\begin{tabular}{c  c c c c c c c c c c} \\ \hline \hline
$L_h$ &  $\beta$  & $T_h$ & C~II & C~II] & C~III] &  C~IV & N~II] & N~III] & 
N~IV] & N~V \\
($\LSUN$) &  & ($10^5$~K)   &$\lambda$1335 &$\lambda$2325 &$\lambda$1908 
&$\lambda$1550 &$\lambda$2141 & $\lambda$1750 & $\lambda$1486 & $\lambda$1240 \\ 
\hline
$6.2\times10^2$ & 0.01 &  0.2 &   -0.8 &   0.3 &  -0.5 &       &  -1.1 &  -1.8 & 
      &       \\ 
$6.2\times10^2$ & 0.01 &  0.5 &   -1.0 &   0.2 &   1.7 &   1.2 &  -1.7 &   0.6 & 
 -0.2 &  -3.3 \\ 
$6.2\times10^2$ & 0.01 &  0.0 &   -0.3 &   0.8 &   2.0 &   2.7 &  -1.4 &   0.7 & 
  1.4 &   1.4 \\ 
$6.2\times10^2$ & 0.01 &  0.0 &    0.3 &   1.2 &   2.0 &   3.0 &  -0.9 &   0.8 & 
  1.4 &   2.1 \\ 
 & & & & & & & & & &  \\
$6.2\times10^3$ & 0.1  &  0.2 &   -1.1 &  -0.1 &  -0.6 &  -6.4 &  -1.5 &  -1.8 & 
 -7.4 &       \\ 
$6.2\times10^3$ & 0.1  &  0.5 &   -1.4 &  -0.4 &   1.3 &   0.9 &  -2.3 &   0.4 & 
 -0.4 &  -2.1 \\ 
$6.2\times10^3$ & 0.1  &  0.0 &   -0.8 &   0.1 &   1.4 &   2.3 &  -2.3 &   0.4 & 
  1.2 &   1.1 \\ 
$6.2\times10^3$ & 0.1  &  0.0 &   -0.3 &   0.6 &   1.4 &   2.7 &  -1.8 &   0.3 & 
  1.1 &   1.7 \\ 
 & & & & & & & & & &  \\
$6.2\times10^4$ & 1    &  0.2 &   -1.4 &  -0.4 &  -1.0 &  -5.9 &  -1.7 &  -2.1 & 
 -7.0 &       \\ 
$6.2\times10^4$ & 1    &  0.5 &   -1.5 &  -0.9 &   0.9 &   0.5 &  -2.8 &   0.3 & 
 -0.5 &  -1.7 \\ 
$6.2\times10^4$ & 1    &  0.0 &   -0.9 &  -0.4 &   1.0 &   1.9 &  -3.0 &   0.3 & 
  1.2 &   0.8 \\ 
$6.2\times10^4$ & 1    &  0.0 &   -0.8 &  -0.2 &   0.8 &   2.4 &  -2.8 &   0.1 & 
  1.1 &   1.3 \\ 
 & & & & & & & & & &  \\
$6.2\times10^5$ & 10   &  0.2 &   -1.5 &  -0.7 &  -1.4 &  -5.4 &  -1.8 &  -2.3 & 
 -6.5 &       \\ 
$6.2\times10^5$ & 10   &  0.5 &   -1.6 &  -1.4 &   0.4 &   0.2 &  -3.2 &   0.1 & 
 -0.6 &  -1.9 \\ 
$6.2\times10^5$ & 10   &  0.0 &   -1.2 &  -1.1 &   0.5 &   1.5 &  -3.6 &   0.2 & 
  0.9 &   0.6 \\ 
$6.2\times10^5$ & 10   &  0.0 &   -0.9 &  -1.0 &   0.4 &   2.0 &  -3.7 &   0.0 & 
  1.1 &   1.0 \\ 
\hline
\end{tabular}
\end{center}


\eject
\normalsize
\centerline{ Table 2. Continued }
\footnotesize
\vspace{-10mm}
\begin{center}
\begin{tabular}{c  c c c c c c c c c c c} \\ \hline \hline
$L_h$ &  $\beta$  & $T_h$ & O~III] &  O~IV] & O~V] & O~V] & O~VI & Mg II & 
Si~II] & Si~III] & Si~IV \\
($\LSUN$) &     & ($10^5$~K) & $\lambda$1664 & $\lambda$1403 & $\lambda$1218 & 
$\lambda$1371& $\lambda$1043& $\lambda$2800& $\lambda$2335& $\lambda$1892& 
$\lambda$1397 \\ \hline
$6.2\times10^2$ & 0.01 &  0.2 &  -2.8 &      &       &       &      &  -1.4 &  
-2.7 & -0.2 &  -3.7 \\
$6.2\times10^2$ & 0.01 &  0.5 &   0.9 & -1.3 &  -3.5 &  -9.9 &      &   0.3 &  
-1.6 &  0.9 &   0.6 \\
$6.2\times10^2$ & 0.01 &  1.0 &   1.7 &  1.4 &   1.6 &  -2.5 &  0.6 &   1.4 &  
-0.7 &  0.8 &   1.0 \\
$6.2\times10^2$ & 0.01 &  2.0 &   1.6 &  1.8 &   2.6 &  -1.3 &  2.9 &   1.6 &  
-0.1 &  1.2 &   1.7 \\
 & & & & & & & & & &  \\
$6.2\times10^3$ & 0.1  &  0.2 &  -2.8 &      &       &       &      &  -2.0 &  
-3.4 & -0.4 &  -3.2 \\
$6.2\times10^3$ & 0.1  &  0.5 &   0.7 & -1.0 &  -2.3 &  -7.4 &      &  -0.1 &  
-2.4 &  0.5 &   0.4 \\
$6.2\times10^3$ & 0.1  &  1.0 &   1.6 &  1.0 &   1.3 &  -2.6 &  0.6 &   1.1 &  
-2.1 &  0.2 &   0.6 \\
$6.2\times10^3$ & 0.1  &  2.0 &   1.5 &  1.4 &   2.4 &  -1.4 &  0.0 &   1.5 &  
-1.2 &  0.7 &   1.2 \\
 & & & & & & & & & &  \\
$6.2\times10^4$ & 1    &  0.2 &  -2.9 & -8.9 &       &       &      &  -2.0 &  
-3.8 & -0.8 &  -2.9 \\
$6.2\times10^4$ & 1    &  0.5 &   0.5 & -1.0 &  -1.8 &  -6.3 & -5.5 &  -0.4 &  
-3.4 &  0.2 &   0.2 \\
$6.2\times10^4$ & 1    &  1.0 &   1.5 &  0.7 &   1.0 &  -2.8 &  0.3 &   1.0 &  
-3.5 & -0.2 &   0.5 \\
$6.2\times10^4$ & 1    &  2.0 &   1.6 &  1.0 &   2.2 &  -1.3 &  2.3 &   1.4 &  
-2.9 &  0.1 &   0.9 \\
 & & & & & & & & & &  \\
$6.2\times10^5$ & 10   &  0.2 &  -3.1 & -8.8 &       &       &      &  -1.5 &  
-3.3 & -1.1 &  -2.9 \\
$6.2\times10^5$ & 10   &  0.5 &   0.2 & -1.3 &  -2.0 &  -6.3 & -5.1 &  -0.8 &  
-3.8 & -0.2 &  -0.1 \\
$6.2\times10^5$ & 10   &  1.0 &   1.3 &  0.6 &   0.7 &  -2.6 & -0.1 &   0.7 &  
-4.2 & -0.8 &   0.3 \\
$6.2\times10^5$ & 10   &  2.0 &   1.6 &  0.7 &   1.9 &  -0.8 &  2.0 &   1.3 &  
-4.4 & -0.4 &   0.7 \\
\hline
\end{tabular}
\end{center}

\noindent
{Note.--Blank spaces mark EWs practically equal to zero.} \\

\noindent
$\rm ^a$ In units of \AA.

\eject

\normalsize
\begin{center}
{ Table 3. Logarithm of model emission line $\rm fluxes^{a}$
for $\MDOT=10^{-7}~\msunyr$, $\gamma=3$ and Vogel's velocity law (V).} 
\footnotesize
{
\begin{tabular}{c  c c c c c c c c c c c c c} 
\\ \hline \hline
$L_h$ &  $\beta$  & $T_h$ & H$\alpha$ & H$\beta$ & P$\alpha$ & He~I & He~I & 
He~I & He~I & He~I & He~I & He~II & He~II \\
$(\LSUN$) &     & ($10^5$~K) & & & &$\lambda$3889 &$\lambda$4471 &$\lambda$5876 
&$\lambda$6678 &$\lambda$7065 &$\lambda$10830 &$\lambda$1640 
& $\lambda$4686 \\ \hline 
$\gamma=3$ & & & & & & & & & & & & & \\ 
$6.2\times10^2$ & 0.01 &  1.0 & -10.9 & -11.7 & -11.8 & -12.3 & -12.4 & -11.9 & 
-12.7 & -12.0 & -10.9 & -11.4 & -12.4 \\
$6.2\times10^3$ & 0.1  &  1.0 &  -9.9 & -10.9 & -10.8 & -11.4 & -11.7 & -10.8 & 
-11.7 & -10.9 & -10.1 & -10.6 & -11.6 \\
$6.2\times10^4$ & 1    &  1.0 &  -9.1 &  -9.7 & -10.1 & -10.0 & -10.7 &  -9.9 & 
-10.6 &  -9.9 &  -9.6 &  -9.6 & -10.6 \\
V & & & & & & & & & & & & & \\ 
$6.2\times10^2$ & 0.01 &  1.0 & -10.7 & -11.6 & -11.7 & -12.2 & -12.3 & -11.8 & 
-12.5 & -11.9 & -10.7 & -11.3 & -12.4 \\
$6.2\times10^3$ & 0.1  &  1.0 &  -9.8 & -10.8 & -10.7 & -11.2 & -11.6 & -10.7 & 
-11.6 & -10.7 & -10.0 & -10.4 & -11.4 \\
$6.2\times10^4$ & 1    &  1.0 &  -9.0 &  -9.6 & -10.0 &  -9.8 & -10.5 &  -9.8 & 
-10.5 &  -9.8 &  -9.5 &  -9.5 & -10.4 \\
$6.2\times10^5$ & 10   &  1.0 &  -8.6 &  -9.0 &  -9.6 &  -9.1 &  -9.6 &  -9.3 &  
-9.4 &  -9.4 &  -9.1 &  -8.5 &  -9.4 \\
\hline
\end{tabular}
}
\end{center}

\eject

\normalsize
\centerline{ Table 3. Continued }
\footnotesize
\vspace{-8mm}
\begin{center}
\begin{tabular}{c  c c c c c c c c c c} \\ \hline \hline
$L_h$ &  $\beta$  & $T_h$ & C~II & C~II] & C~III] &  C~IV & N~II] & N~III] & 
N~IV] & N~V \\
($\LSUN$) &  & ($10^5$~K)   &$\lambda$1335 &$\lambda$2325 &$\lambda$1908 
&$\lambda$1550 &$\lambda$2141 & $\lambda$1750 & $\lambda$1486 & $\lambda$1240 \\ 
\hline
$\gamma=3$ & & & & & & & & & &  \\
$6.2\times10^2$ & 0.01 &  1.0 & -12.3 & -11.9 & -10.6  & -9.9 & -13.8 & -11.6 & 
-11.0 & -11.0 \\
$6.2\times10^3$ & 0.1  &  1.0 & -11.7 & -11.6 & -10.0  & -9.0 & -13.7 & -10.8 & 
-10.0 & -10.2 \\
$6.2\times10^4$ & 1    &  1.0 & -11.0 & -11.3 &  -9.5  & -8.3 & -13.6 &  -9.9 &  
-9.0 &  -9.3 \\
V & & & & & & & & & &  \\
$6.2\times10^2$ & 0.01 &  1.0 & -12.2 & -11.9 & -10.5  & -9.8 & -13.8 & -11.5 & 
-10.9 & -11.0 \\
$6.2\times10^3$ & 0.1  &  1.0 & -11.6 & -11.5 & -10.0  & -8.9 & -13.6 & -10.7 &  
-9.9 & -10.1 \\
$6.2\times10^4$ & 1    &  1.0 & -10.8 & -11.1 &  -9.4  & -8.1 & -13.3 &  -9.7 &  
-8.8 &  -9.2 \\
$6.2\times10^5$ & 10   &  1.0 &  -9.9 & -10.8 &  -8.8  & -7.5 & -12.9 &  -8.8 &  
-8.0 &  -8.2 \\
\hline
\end{tabular}
\end{center}


\eject
\normalsize
\centerline{ Table 3. Continued }
\footnotesize
\vspace{-10mm}
\begin{center}
\begin{tabular}{c  c c c c c c c c c c c} \\ \hline \hline
$L_h$ &  $\beta$  & $T_h$ & O~III] &  O~IV] & O~V] & O~V] & O~VI & Mg II & 
Si~II] & Si~III] & Si~IV \\
($\LSUN$) &     & ($10^5$~K) & $\lambda$1664 & $\lambda$1403 & $\lambda$1218 & 
$\lambda$1371& $\lambda$1034 & $\lambda$2800& $\lambda$2335& $\lambda$1892& 
$\lambda$1397 \\ \hline
$\gamma=3$ & & & & & & & & & & &  \\
$6.2\times10^2$ & 0.01 &  1.0 & -10.7 & -11.0 & -10.7 & -15.1 & -11.3 & -11.3 & 
-13.3 & -11.6 & -11.2 \\
$6.2\times10^3$ & 0.1  &  1.0 &  -9.7 & -10.2 &  -9.9 & -14.2 & -10.5 & -10.7 & 
-13.7 & -11.1 & -10.4 \\
$6.2\times10^4$ & 10   &  1.0 &  -8.7 &  -9.2 &  -8.9 & -13.1 &  -9.7 &  -9.9 & 
-14.4 & -10.6 &  -9.5 \\
V & & & & & & & & & & &  \\
$6.2\times10^2$ & 0.01 &  1.0 & -10.6 & -11.0 & -10.8 & -15.3 & -11.4 & -11.2 & 
-13.4 & -11.5 & -11.1 \\
$6.2\times10^3$ & 0.1  &  1.0 &  -9.6 & -10.1 &  -9.8 & -14.2 & -10.5 & -10.5 & 
-13.6 & -11.0 & -10.3 \\
$6.2\times10^4$ & 1    &  1.0 &  -8.6 &  -9.1 &  -8.8 & -13.0 &  -9.6 &  -9.7 & 
-14.1 & -10.4 &  -9.4 \\
$6.2\times10^5$ & 10   &  1.0 &  -7.8 &  -8.2 &  -7.9 & -11.4 &  -8.8 &  -9.1 & 
-14.7 & -10.1 &  -8.5 \\
\hline
\end{tabular}
\end{center}

\noindent
{Note.--Blank spaces mark fluxes practically equal to zero.}\\

\noindent
$\rm ^a$ {In units of erg~$\rm  cm^{-2}~s^{-1}$.} 


\begin{figure}[htbp]
\epsfxsize=6.0in
\epsffile{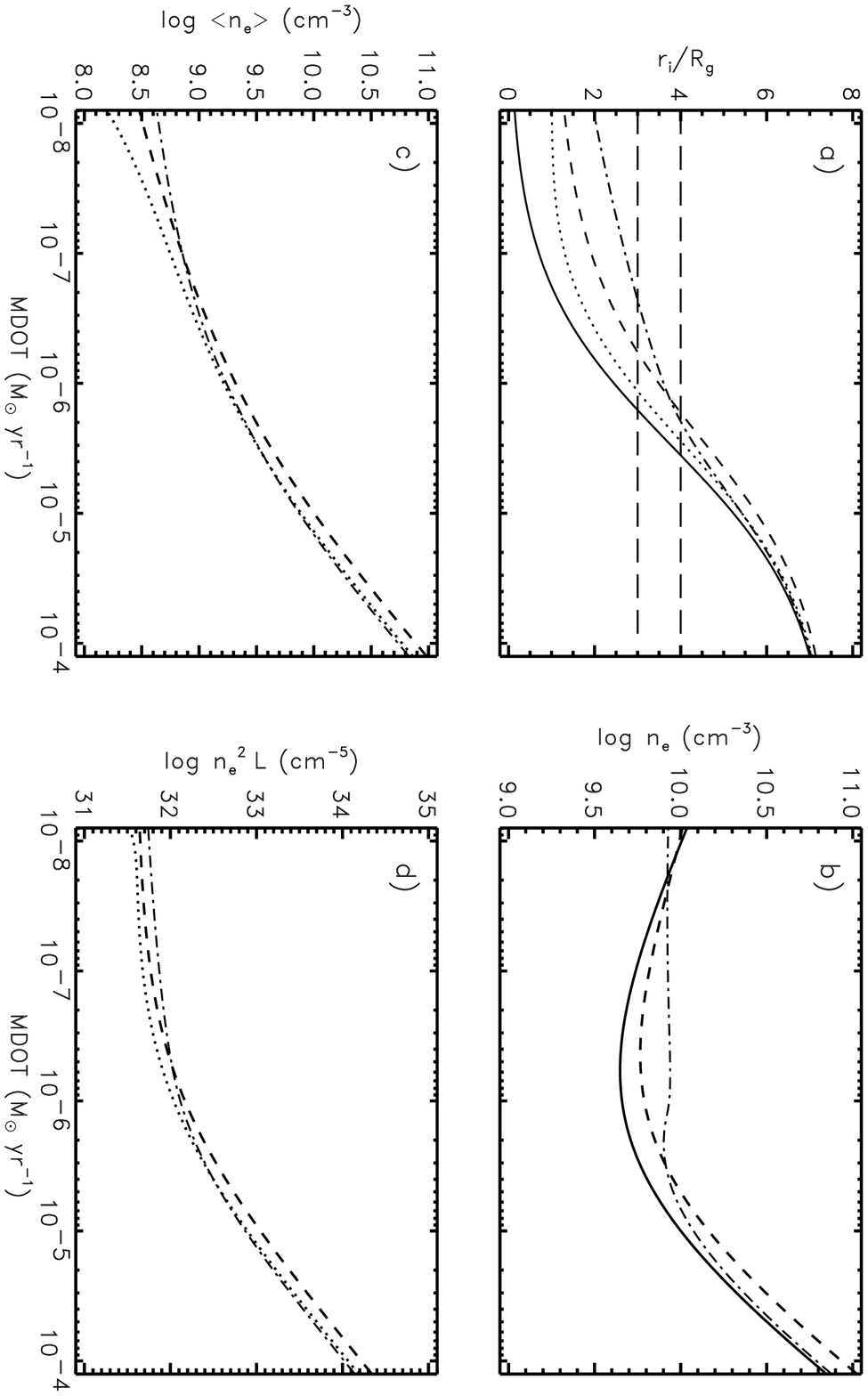}
\caption {}
\end{figure}

\begin{figure}[htbp]
\epsfxsize=6.0in
\epsffile{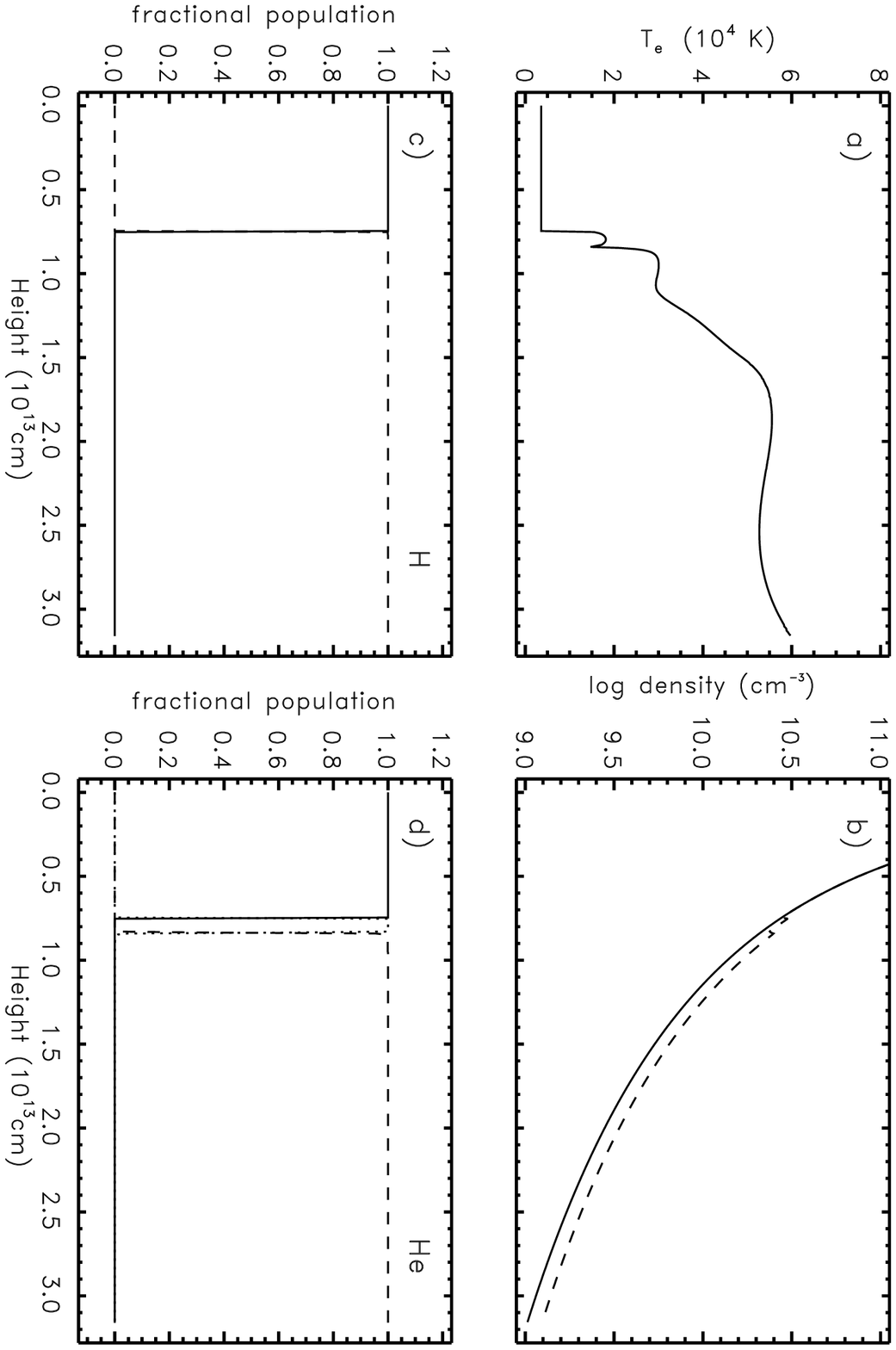}
\caption {}
\end{figure}

\begin{figure}[htbp]
\epsfxsize=6.0in
\epsffile{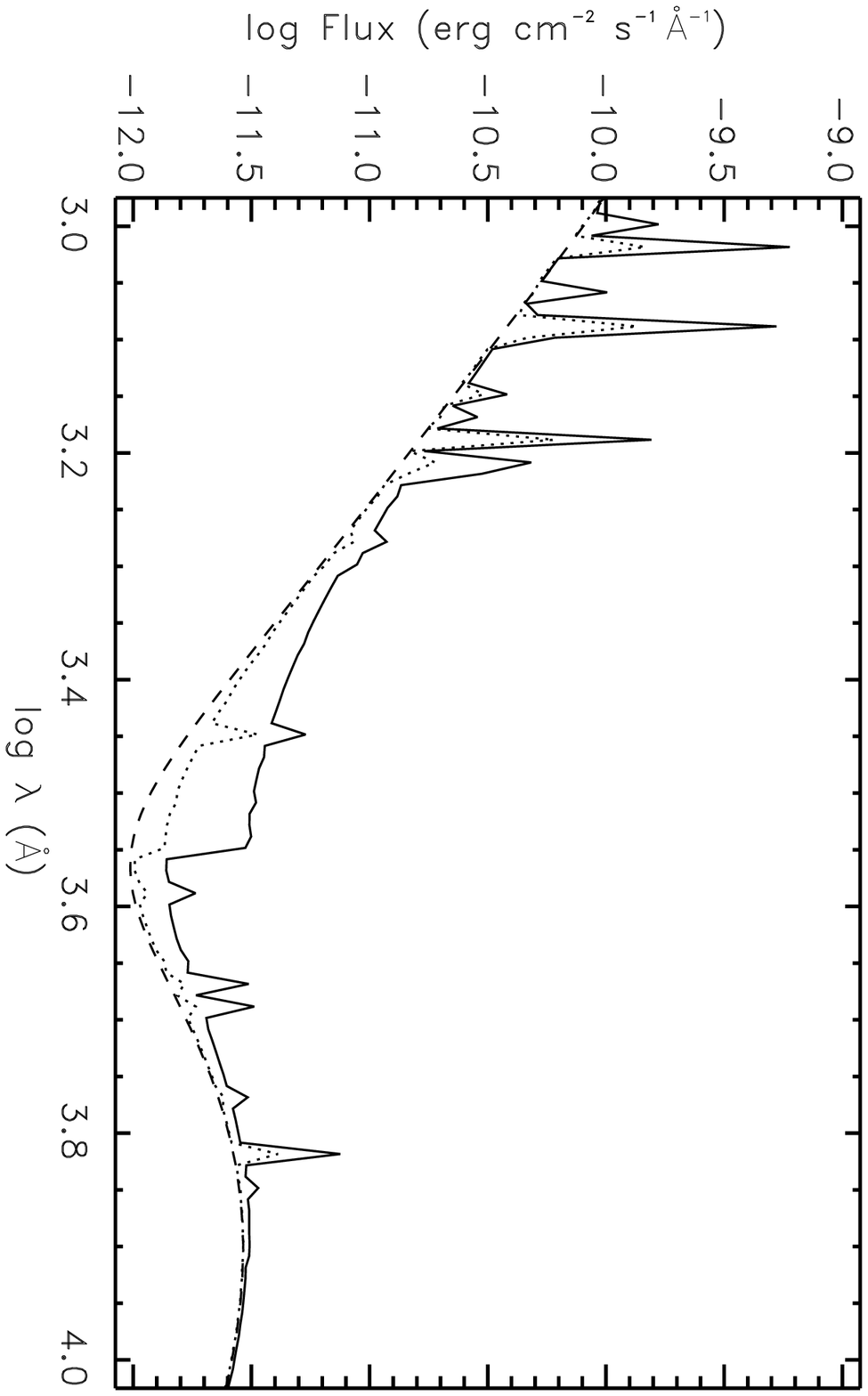}
\caption {}
\end{figure}

\epsfxsize=6.0in
\epsffile{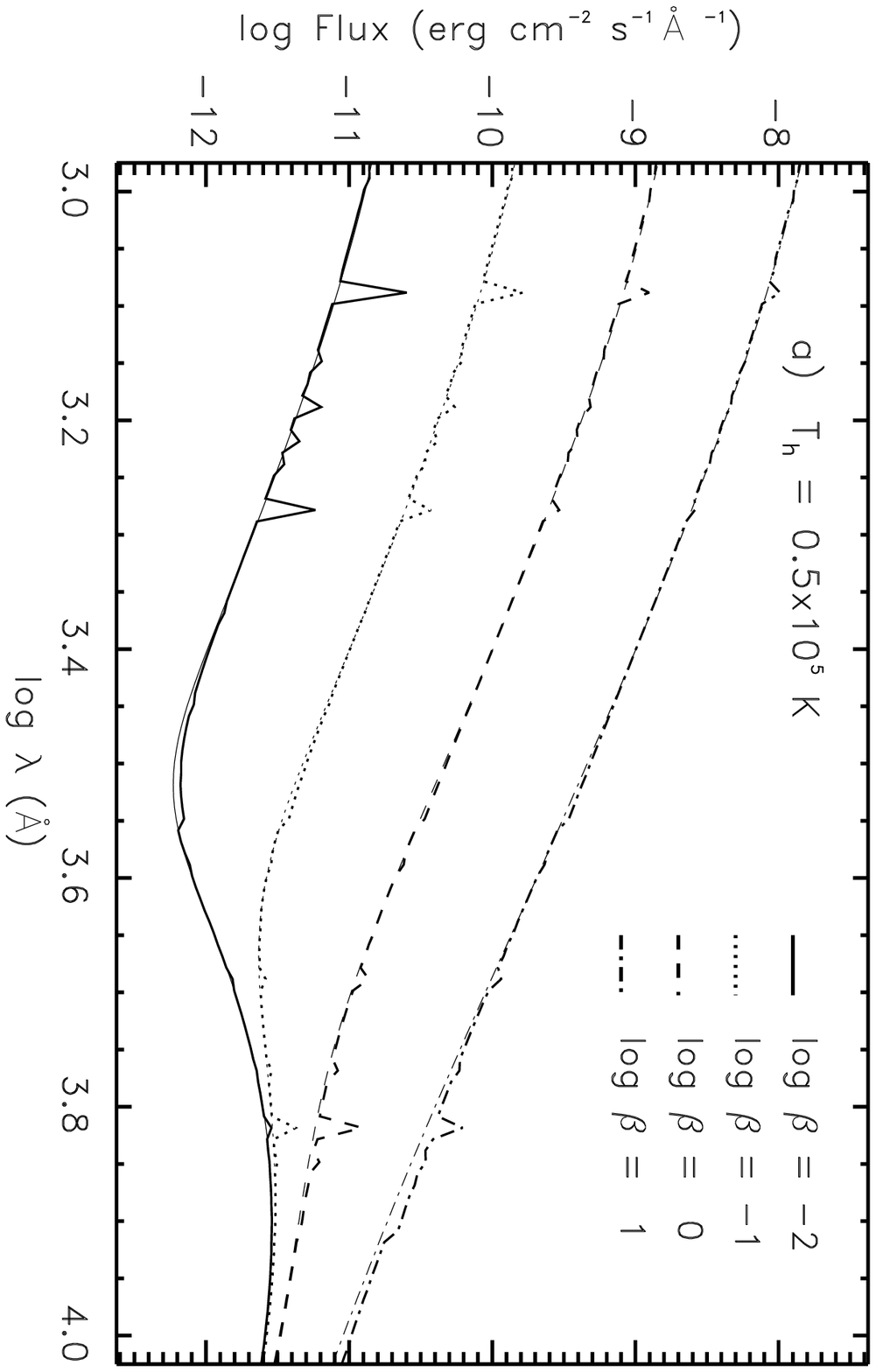}

\epsfxsize=6.0in
\epsffile{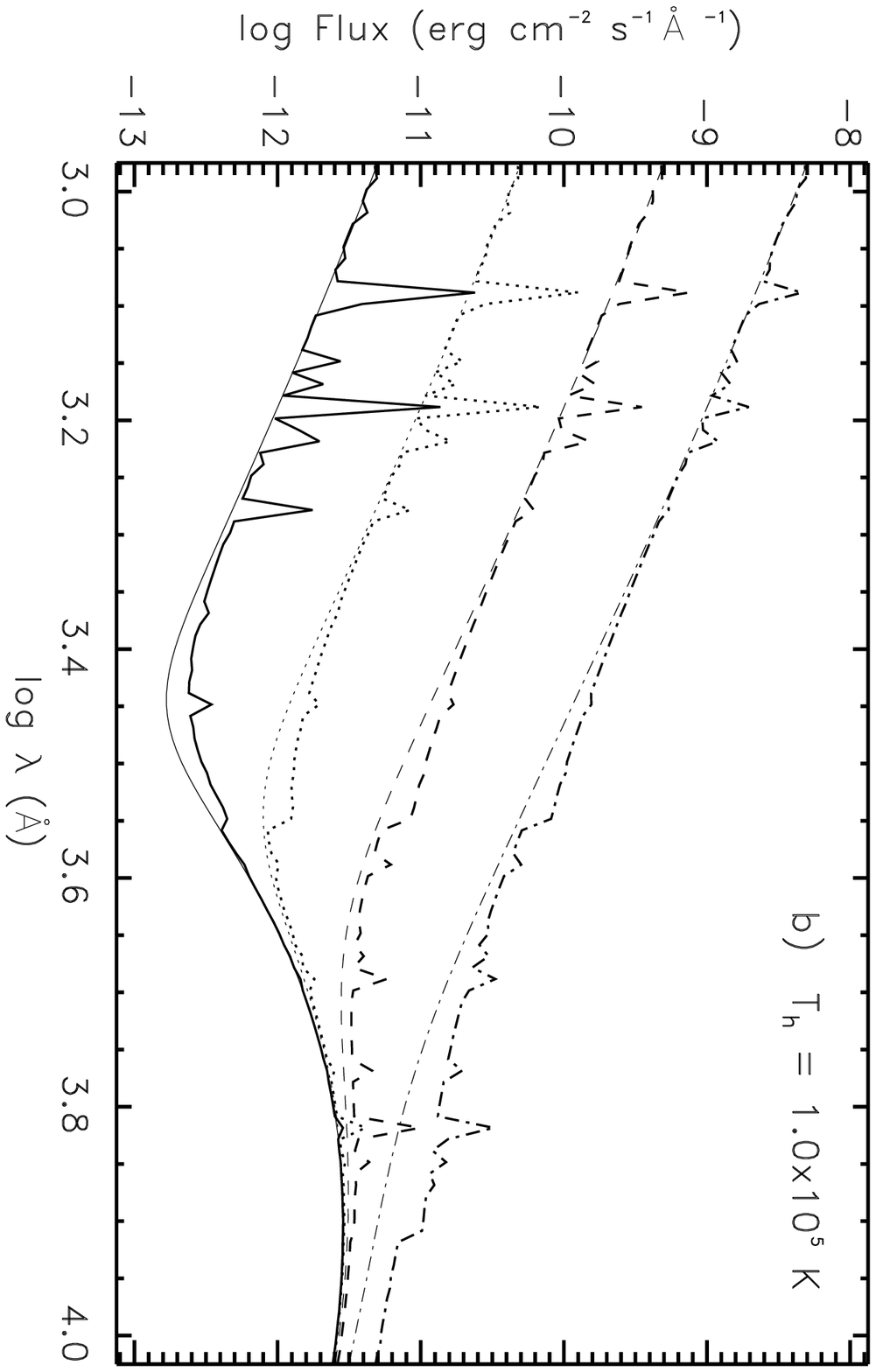}

\begin{figure}[htbp]
\epsfxsize=6.0in
\epsffile{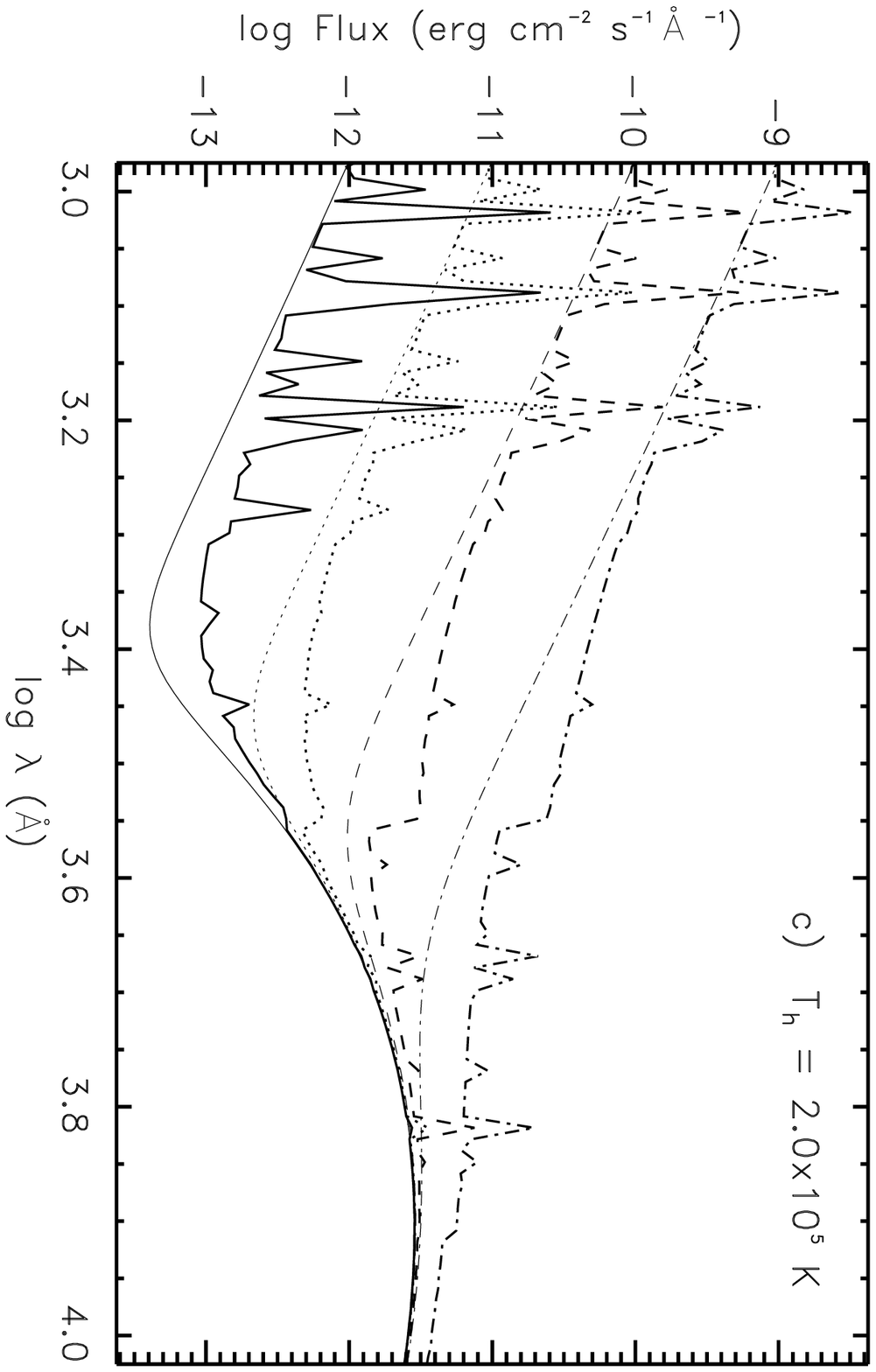}
\caption {} 
\end{figure}

\begin{figure}[htbp]
\epsfxsize=6.0in
\epsffile{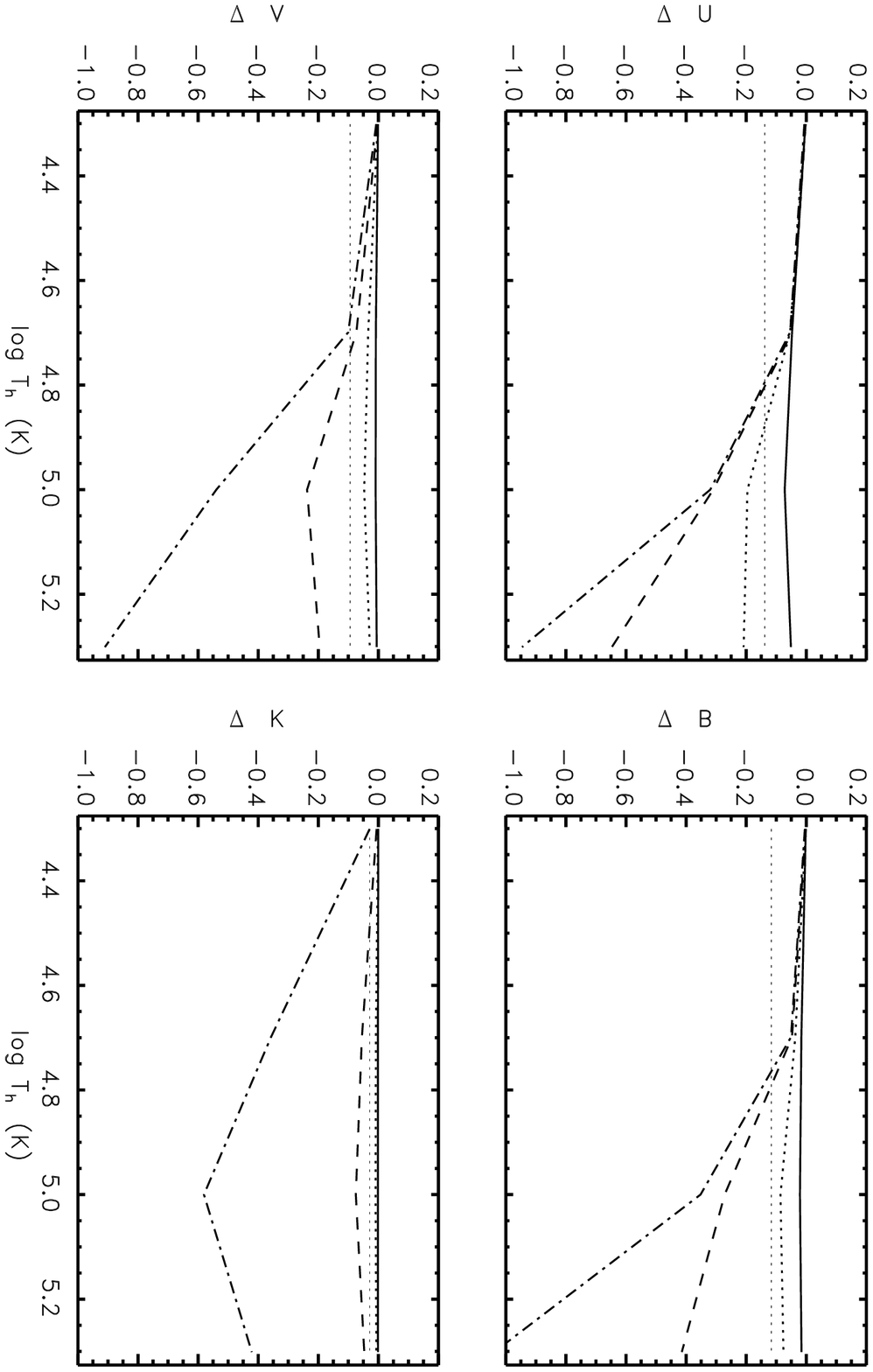}
\caption {} 
\end{figure}

\begin{figure}[htbp]
\epsfxsize=6.0in
\epsffile{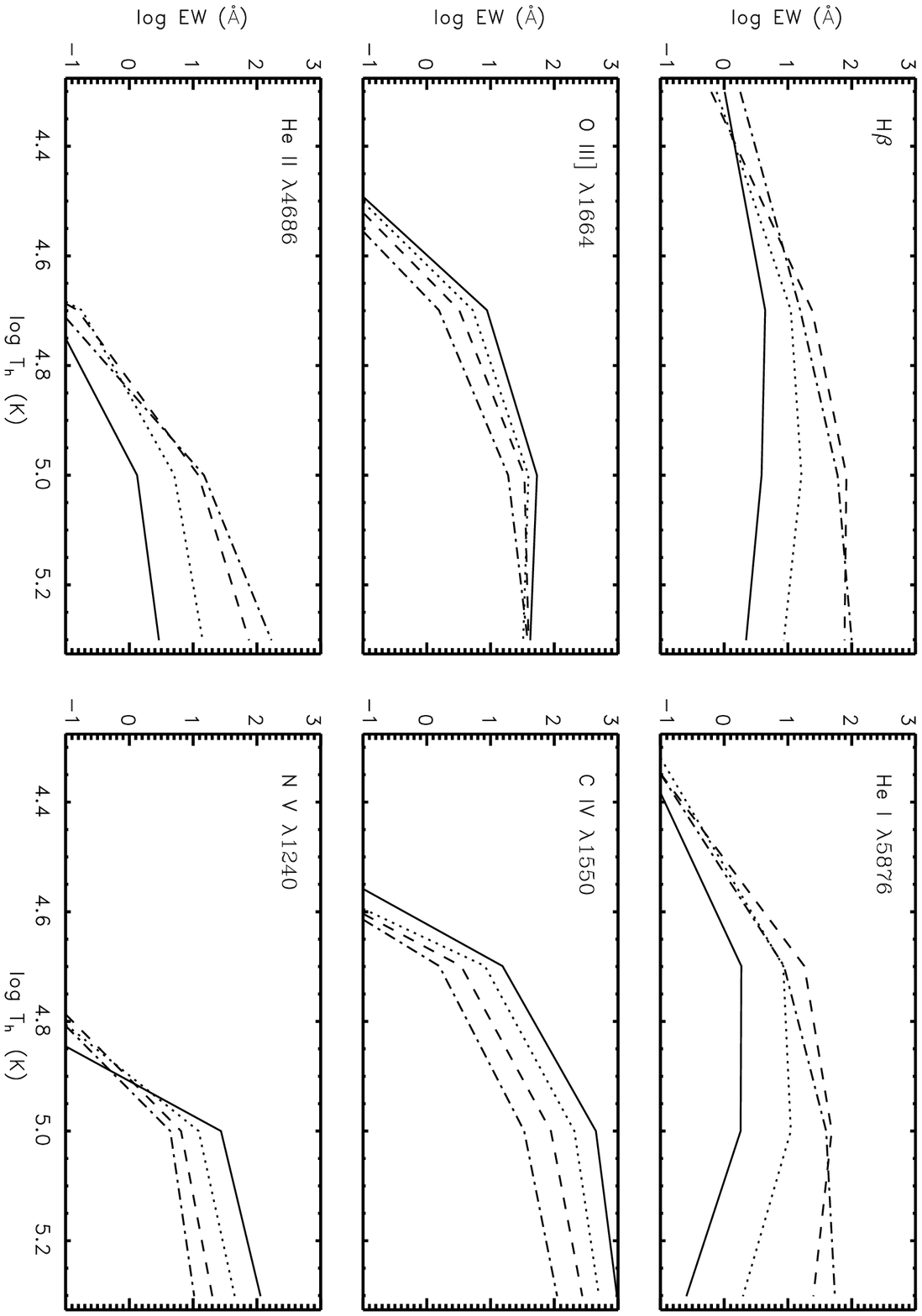}
\caption {} 
\end{figure}

\begin{figure}[htbp]
\epsfxsize=6.0in
\epsffile{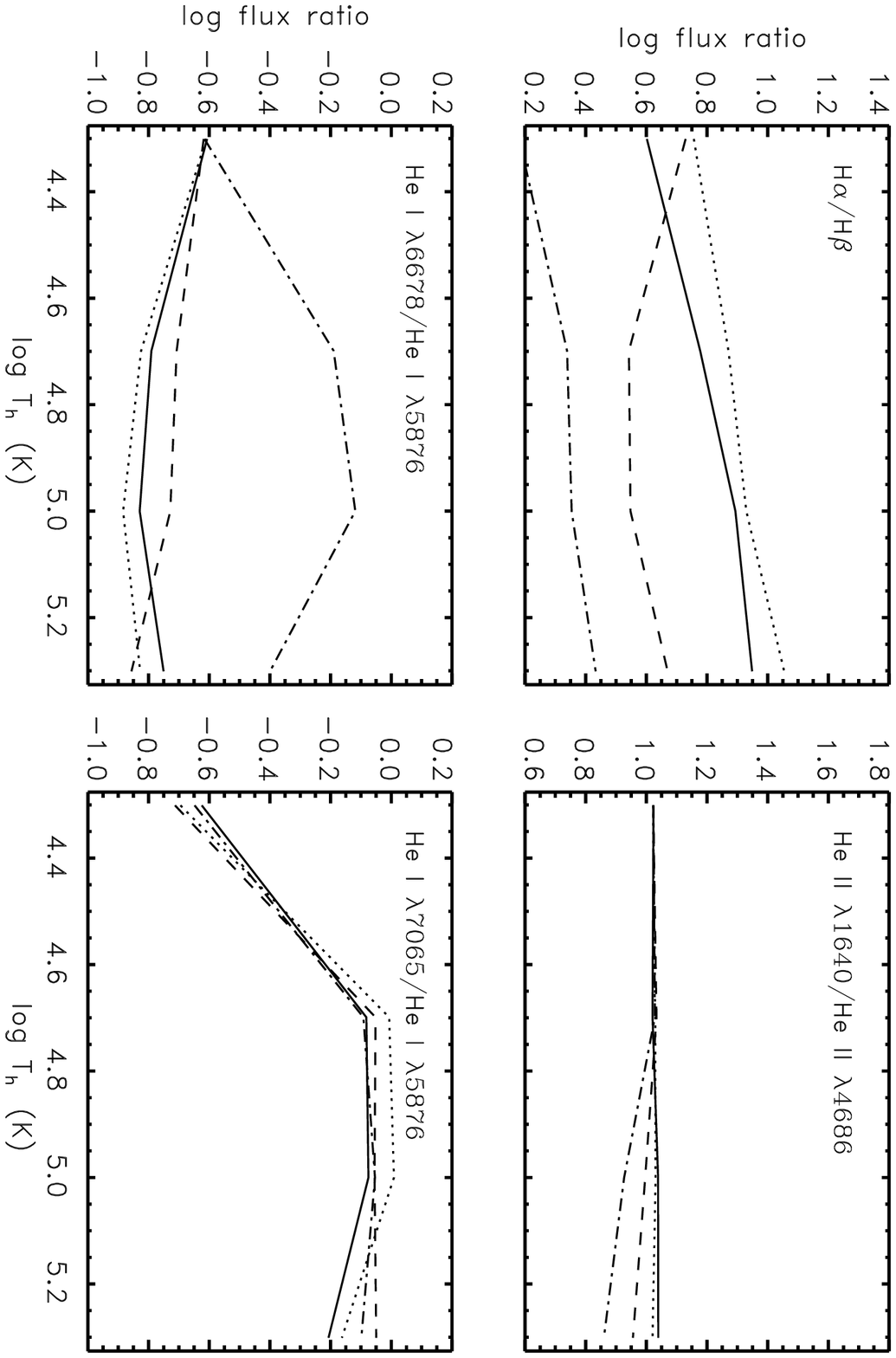}
\caption {} 
\end{figure}

\begin{figure}[htbp]
\epsfxsize=6.0in
\epsffile{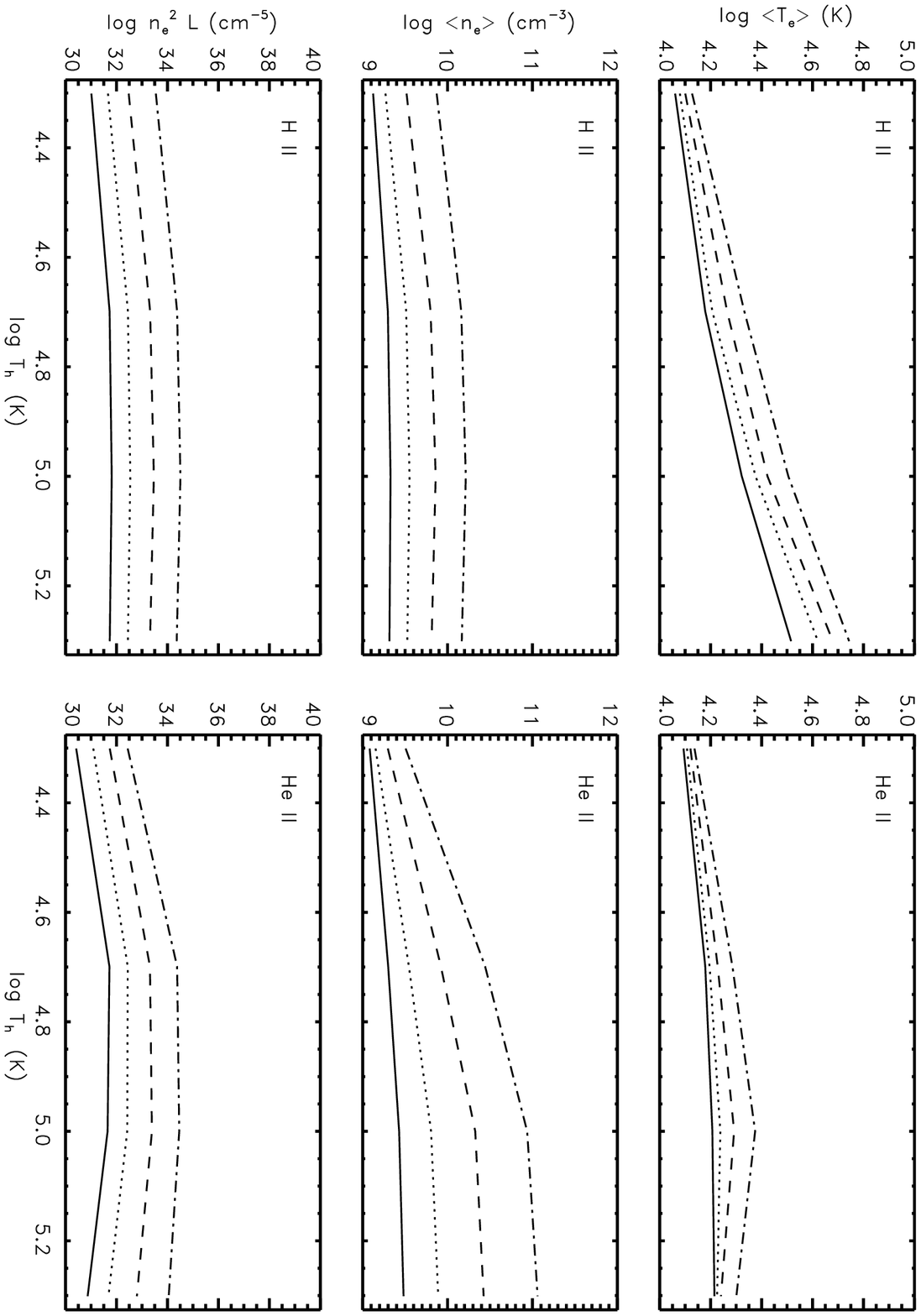}
\caption {} 
\end{figure}

\begin{figure}[htbp]
\epsfxsize=6.0in
\epsffile{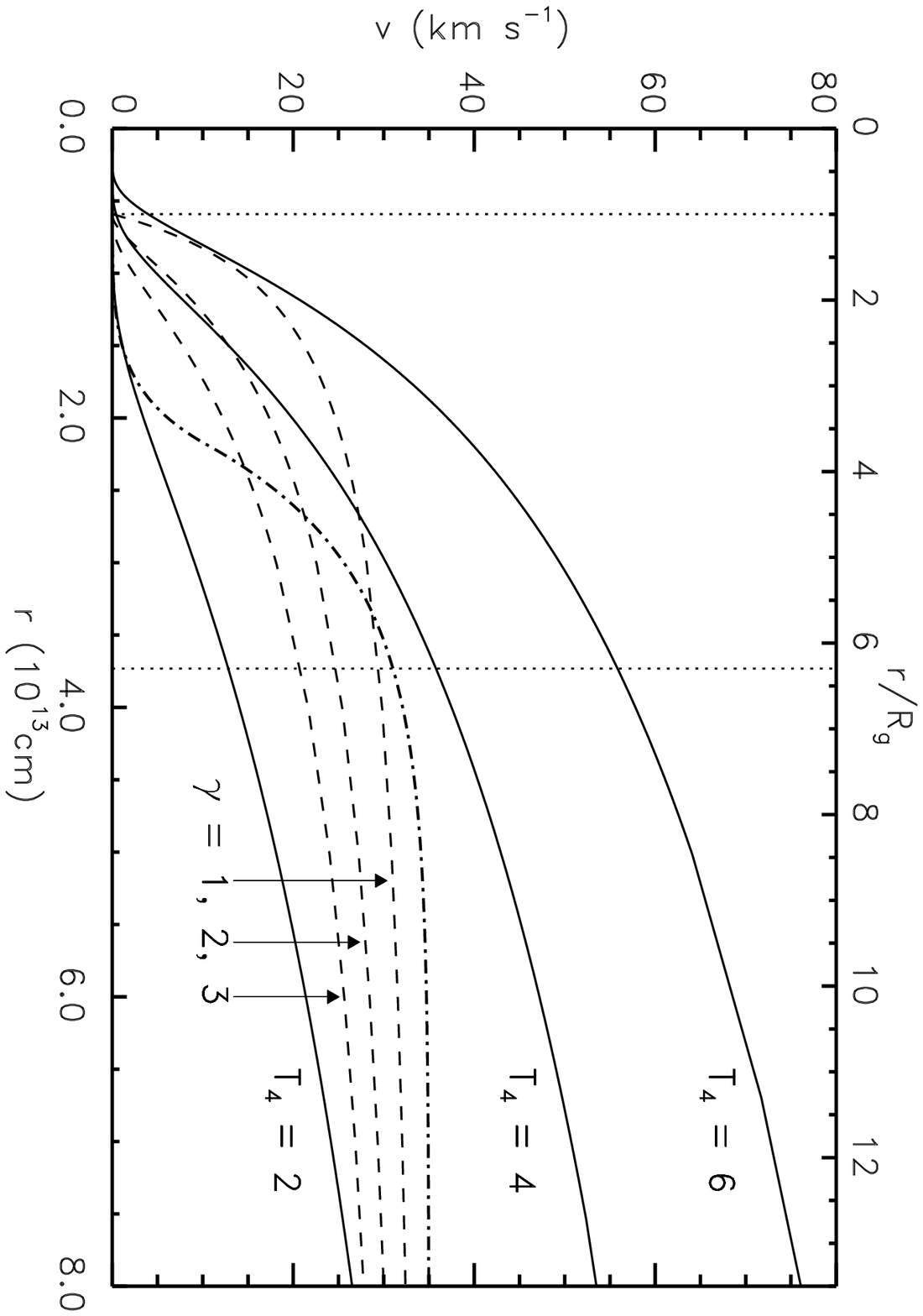}
\caption {} 
\end{figure}

\begin{figure}[htbp]
\epsfxsize=6.0in
\epsffile{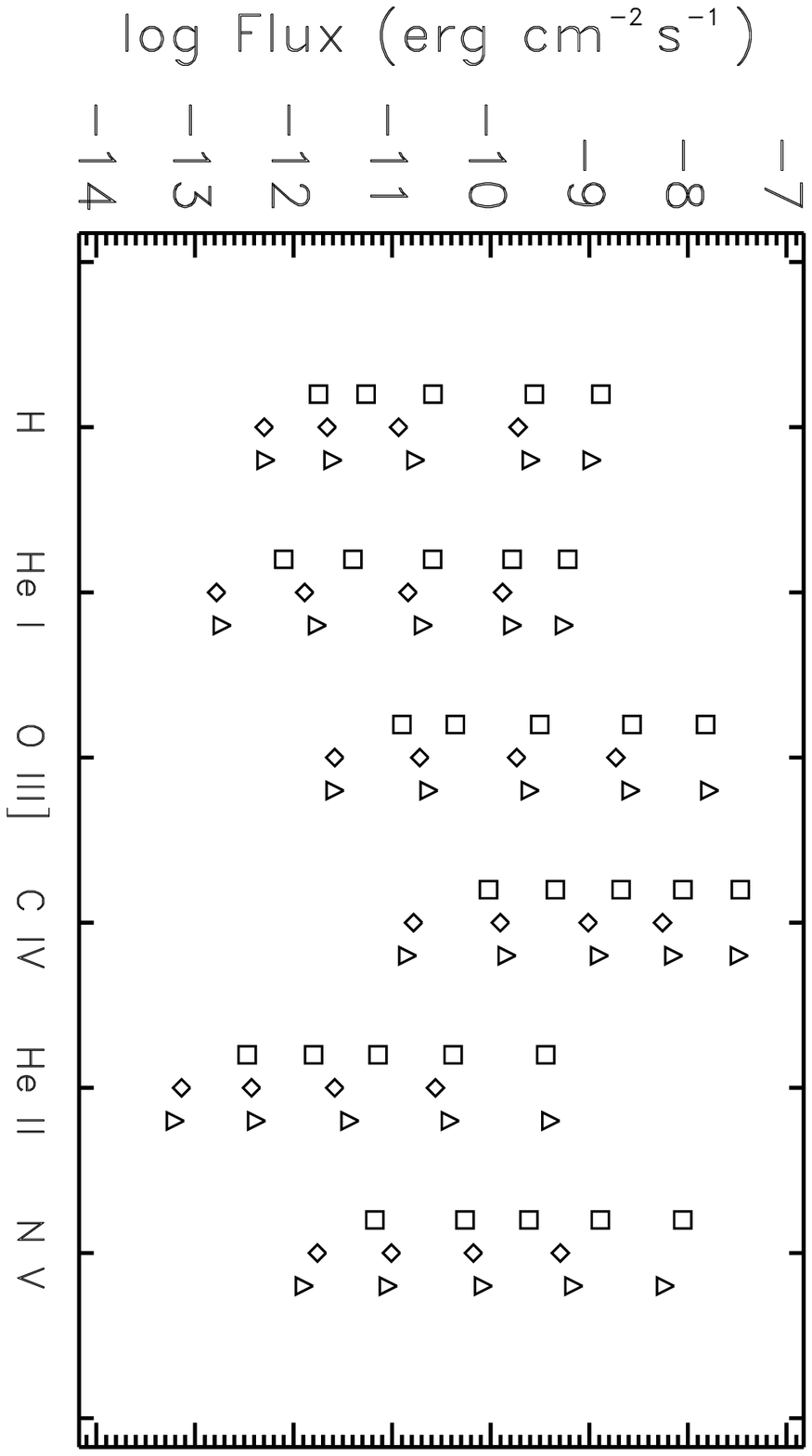}
\caption {} 
\end{figure}

\begin{figure}[htbp]
\epsfxsize=6.0in
\epsffile{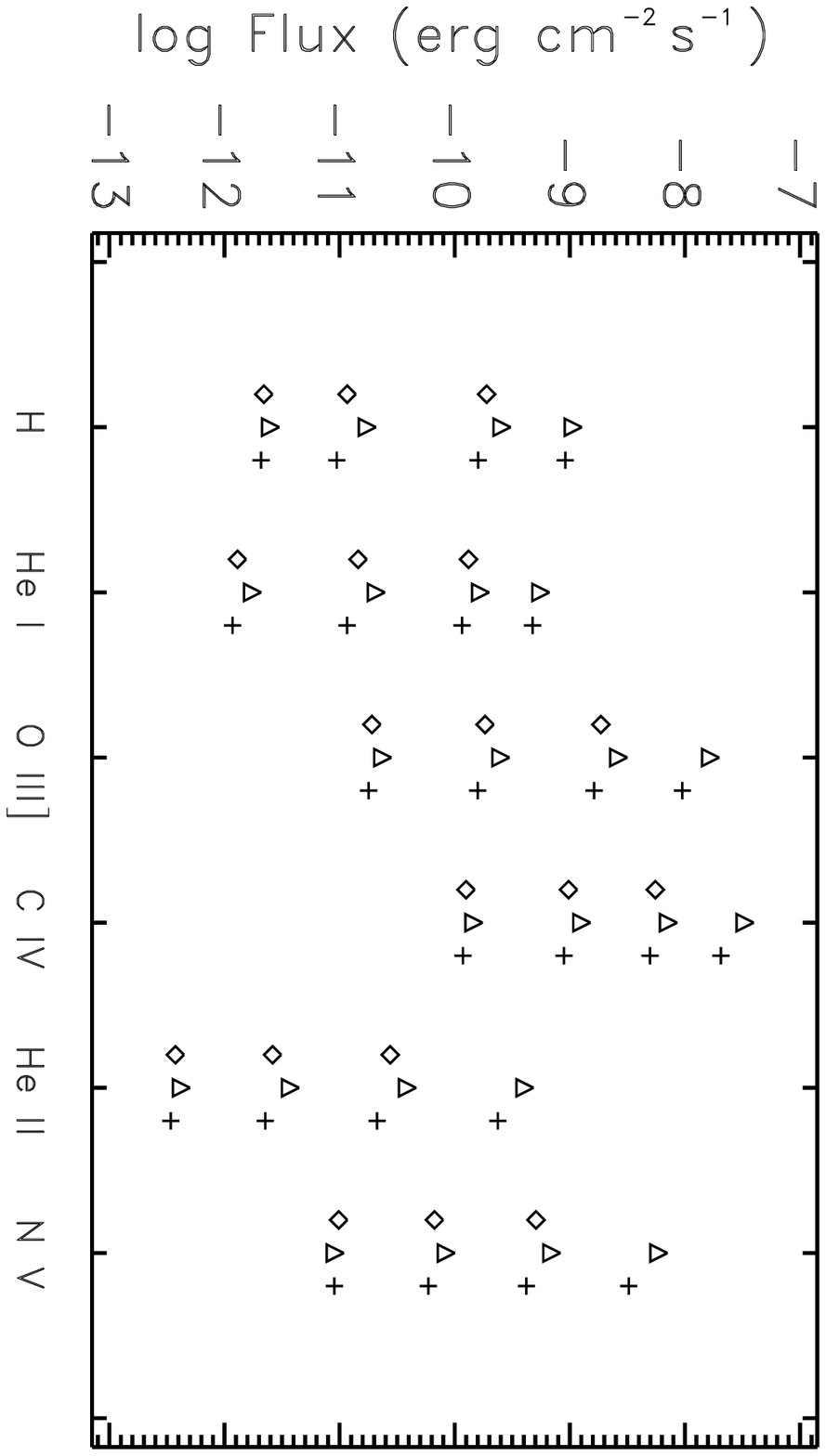}
\caption {} 
\end{figure}

\begin{figure}[htbp]
\epsfxsize=6.0in
\epsffile{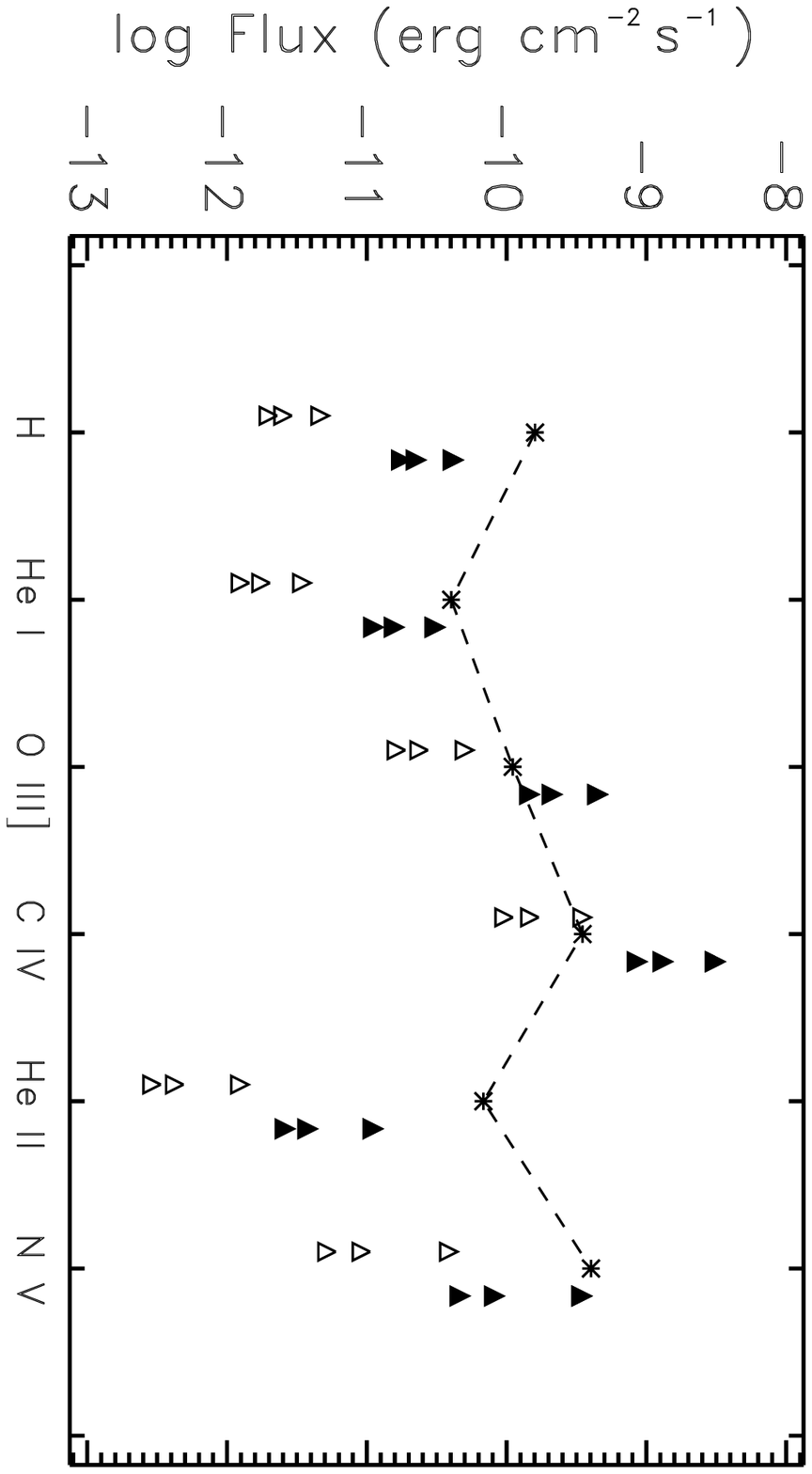}
\caption {} 
\end{figure}

\begin{figure}[htbp]
\epsfxsize=6.0in
\epsffile{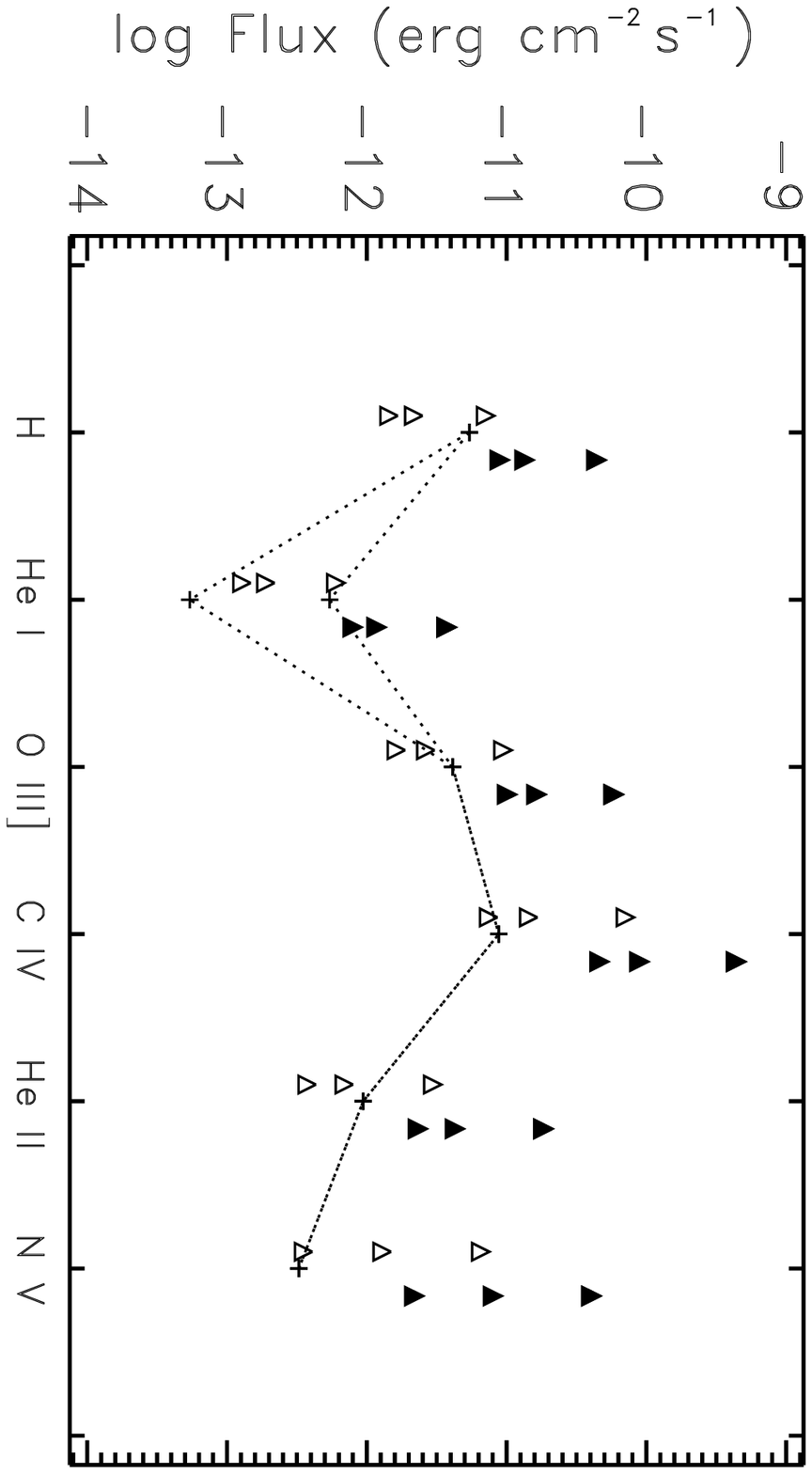}
\caption {} 
\end{figure}

\end{document}